\documentclass[a4paper,10pt]{article}

\usepackage{a4wide}
\usepackage{amsmath}
\usepackage{amssymb}
\usepackage{epsfig}
\usepackage{multirow}
\usepackage{subfigure}
\usepackage{url}
\usepackage{verbatim} 

\begin{document}

\title{Capacity Planning for Vertical Search Engines}

\newcommand{\tabauthors}{
\begin{tabular}{ccc}
Claudine Badue$^1$
& Jussara Almeida$^2$
& Virg\'ilio Almeida$^2$
\end{tabular}\\ 
\begin{tabular}{cc}
Ricardo Baeza-Yates$^3$
& Berthier Ribeiro-Neto$^2~^4$ 
\end{tabular}\\
\begin{tabular}{cc}
Artur Ziviani$^5$
& Nivio Ziviani$^2$
\end{tabular}
}

\newcommand{\tabinstitutions}{
\begin{tabular}{c}
$^1$Federal University of Esp\'irito Santo, Vit\'oria, Brazil\\                      
\url{claudine@lcad.inf.ufes.br}        
\end{tabular}\\ [2ex]
\begin{tabular}{c}
$^2$Federal University of Minas Gerais, Belo Horizonte, Brazil\\
\mbox{\url{{berthier,jussara,virgilio,nivio}@dcc.ufmg.br}}
\end{tabular}\\[2ex]
\begin{tabular}{c}
$^3$Yahoo! Research, Barcelona, Spain \& Santiago, Chile\\
\url{ricardo@baeza.cl}
\end{tabular}\\[2ex]
\begin{tabular}{c}
$^4$Google Engineering, Belo Horizonte, Brazil\\
\end{tabular}\\[2ex]
\begin{tabular}{c}
$^5$National Laboratory for Scientific Computing (LNCC), Petr\'opolis, Brazil \\
\url{ziviani@lncc.br}
\end{tabular}
}

\author{\tabauthors \\[4ex] \tabinstitutions}

\date{}

\maketitle

\date{}

\maketitle

\begin{abstract}
Vertical search engines focus on specific slices of content, such as the Web of a single country or the document collection of a large corporation. Despite this, like general open web search engines, they are expensive to maintain, expensive to operate, and hard to design. Because of this, predicting the response time of a vertical search engine is usually done empirically through experimentation, requiring a costly setup. An alternative is to develop a model of the search engine for predicting performance. However, this alternative is of interest only if its predictions are accurate. In this paper we propose a methodology for analyzing the performance of vertical search engines. Applying the proposed methodology, we present a capacity planning model based on a queueing network for search engines with a scale typically suitable for the needs of large corporations. The model is simple and yet reasonably accurate and, in contrast to previous work, considers the imbalance in query service times among homogeneous index servers. We discuss how we tune up the model and how we apply it to predict  the impact on the query response time when parameters such as CPU and disk capacities are changed. This allows a manager of a vertical search engine to determine a priori whether a new configuration of the system might keep the query response under specified performance constraints.\\[2ex]
\end{abstract}

\section{Introduction}



Vertical search engines are used to search collections 
composed of documents of a large corporation or a subset of the Web, such as
the Web of a single country or a set of documents related to
a specific domain (e.g. medical information). 
They require a large number of computational resources to handle
the incoming query traffic, which is often characterized by high peak
requirements on query load.
To cope with these requirements, modern vertical search engines might rely on large clusters of server machines for query processing, a configuration similar to
the one observed for general Web search engines~\cite{Brewer01,Barroso03,Risvik03}, although at a smaller scale.

In line with this trend, we consider a search engine architecture composed of a cluster of index servers, with the documents partitioned among them so that each index server stores a part of the document collection and an index for it.
This architecture is usually referred to as {\em document partitioning}~\cite{Tomasic93,Ribeiro-Neto98}.
The document partitioning architecture is preferred because it simplifies
maintenance, simplifies the generation of the index,
which can be done locally, and degrades gracefully because the failure of an
index server does not prevent any query from being answered, though the final
answer set might not contain all the relevant documents in the collection.
The cluster also includes a broker that communicates with the various index servers.
A new user query reaches the search engine through the broker, which sends a copy of
the query to each index server for local processing.
Afterwards, the broker receives the top ranked documents from each index server and runs a merge to determine the final set of answers to be sent to the user.

The processing of a query is split into two consecutive major phases~\cite{Barroso03}.
The first phase retrieves references to the documents that contain all query terms
and generates a ranking according to a relevance metric, a task usually done
by the index servers.
The second phase takes the merged top ranked answers 
and generates snippets, title, and pointer information for each of them.
This task is usually performed by a cluster of document servers, each one holding a part of the document collection. While the second phase has a roughly constant cost, independent of the size of the document collection, the first phase has a cost that increases with the size of the collection. Therefore, the performance of the first phase is crucial for maintaining the scalability of modern search engines that deal with an ever-increasing number of available documents.

In this paper we propose a capacity planning methodology for vertical search engines
based on estimates of the average response time for retrieving the most relevant
documents for a given user query, i.e., the first and typically most costly phase of the query processing task.
Our methodology is composed of two major steps: (i)~the {\it characterization of real vertical search engine workloads}, and (ii)~the {\it performance modeling} of a vertical search engine architecture itself. The characterization of real workloads is a key step in any capacity planning task, as it allows one to uncover key properties 
of the workload~\cite{Menasce04}. 
Here, we have analyzed and experimented with query logs from two different real-world vertical search engines, characterizing the distributions of query and term popularities, number of terms per query, query interarrival times, and query service times.

The heart of our methodology is a performance model which, in contrast to previous work (see Section~\ref{section:relatedwork} for further details), captures the impact of the {\it imbalance} in query service times among homogeneous index servers. As will be shown, this imbalance can severely degrade the average query system response time. Thus, our model provides much more accurate performance estimates than the existing literature.
Moreover, driven by the assumptions of exponential query interarrival times and service times, observed in  {\it real} workloads, our  performance model is simple (but not simplistic) and relies only on data that are easy to collect using standard tools, thus being of practical use for cost-effective capacity planning decisions in real-world setups.

We demonstrate the applicability of our performance model
in a number of illustrative scenarios.
Once the key parameters have been estimated from easy-to-collect, small scale experimental data,
our performance model can be used to investigate the behavior, provide upper-bound estimates, and ultimately drive the capacity planning decisions  of a large vertical search engine.
For instance, we consider a collection of $1$~billion documents partitioned among $100$~index servers and analyze the impact on the query system response time of adopting larger main memories, faster CPUs and disks as well as adopting application-level caching of query results at the broker.
To the best of our knowledge, this is the first work to propose a 
methodology for capacity planning of vertical search engines.

This paper is organized as follows.
Section~\ref{section:relatedwork} discusses research related to ours, while the
architecture of our target  search engine is presented in Section~\ref{section:architecture}.
Our capacity planning methodology for modern vertical search engine architectures is introduced in the following two sections: the characterization
of key properties of real search engine workloads is presented in Section~\ref{section:workload}, and  the design and validation of our performance model is given in Section~\ref{section:capacityplanning}. Section~\ref{section:applicability} shows, step-by-step, how our model can be applied in a practical capacity planning case study. Finally, conclusions and possible directions of future work are
offered in Section~\ref{section:conclusions}.

\section{Related Work}
\label{section:relatedwork}

In this section, we discuss related work in three key areas to capacity planning for
vertical search engines: index organization in Section~\ref{section:relatedwork:index},
workload characterization in Section~\ref{section:relatedwork:workload}, and
performance modeling in Section~\ref{section:relatedwork:performance}.

\subsection{Index Organization}
\label{section:relatedwork:index}

Different strategies for distributing the index of a document collection among machines have been discussed in literature.
Tomasic and Garcia-Molina~\cite{Tomasic93} compare the performance impact on query processing of two basic and distinct options for storing the inverted lists, namely document partitioning and term partitioning.
In document partitioning, documents are distributed among a set of index servers and each server stores a full index for its subset of the documents.
In term partitioning, the index is distributed across the set of index servers and each server stores full index information for a subset of the terms.
Simulation experiments attempt to determine under which conditions each index organization is better, how each index organization  scales up to large systems and what is the impact of key parameters, such as seeking time of the storage device, load level, and number of keywords in a ``boolean and'' query.

Jeong and Omiecinski~\cite{Jeong95} consider the document partitioning  and the term partitioning approaches to physically divide inverted indexes in a shared-everything multiprocessor machine with multiple disks.
By simulation, they study the performance impact of these schemes on boolean query processing under a number of workloads where the term frequencies in the documents, the term frequencies in the queries, the number of disks, and the multiprogramming level vary.
Ribeiro-Neto and Barbosa~\cite{Ribeiro-Neto98} study how the performance of disjunctive queries is affected by the index organization, the network capacity, and the disk transfer rates, using a simple analytical model coupled with a small simulator.
MacFarlane et al.~\cite{MacFarlane00} and Badue et al.~\cite{Badue01} investigate the performance impact on parallel query processing of the two distinct types of index organizations (either document or term partitioning), using a real case implementation.
Sornil and Fox~\cite{Sornil01} and Badue et al.~\cite{Badue02} compare the performance of hybrid partitioning against document partitioning and term partitioning.
In the hybrid scheme they divide an inverted list into a number of equal sized chunks, which are randomly distributed to nodes in the system.


The previous comparisons of different strategies for distributing the index of a document collection have reached conclusions that are dependent on the use case considered. Whereas Tomasic and Garcia-Molina~\cite{Tomasic93} and Sornil and Fox~\cite{Jeong95}  find in favor of document partitioning, term partitioning has been pointed out as a better alternative by  Ribeiro-Neto and Barbosa~\cite{Ribeiro-Neto98}  and Badue et al.~\cite{Badue01}. Further, Sornil and Fox~\cite{Sornil01} and Badue et al.~\cite{Badue02}  show that the hybrid partitioning outperforms both document partitioning and term partitioning.  Regardless of the conclusion,
these studies suffer from limitations such as the use of artificial or small sets of documents or queries, which are unlikely to be predictive of real-world behavior, and the use of (perhaps too simple and thus unrealistic) simulation models to estimate response time.

Under conditions better approximating those of a real-world large-scale search engine, Moffat et al.~\cite{Moffat06,Moffat07} introduce a pipelined query evaluation methodology based on term partitioning, in which partially evaluated queries are passed amongst the set of index servers that host the query terms.
Moffat et al.~\cite{Moffat06} examine methods for load balancing in the pipelined query evaluation methodology based on term partitioning and propose a suite of techniques for reducing net querying costs.
In particular, they explore the load distribution behavior that pipelining displays, and show that the imbalances can be addressed by techniques that include predictive index list assignments to nodes and selective index list replication.
An important conclusion of their experimental investigation is that document partitioning retains its leading position as the method against which others must be judged.

We thus adopt  document partitioning for distributing the index of our document collection among index servers
(see Section~\ref{section:architecture}) because of its superior performance and popularity in large-scale search engines~cite{Barroso03,Risvik03}.

\subsection{Workload Characterization}
\label{section:relatedwork:workload}

Previous work on query characterization for Web search engines mainly focuses on the characterization of user search behavior and user search goals to enhance the relevance to users of the provided answers, i.e., to improve the search \textit{efficacy}.
Silverstein et al.~\cite{Silverstein99} present an analysis of individual queries, query duplication, and query sections in a query log from the AltaVista Search Engine.
Spink et al.~\cite{Spink00} examine the query reformulation by users, and particularly the use of relevance feedback by users of the Excite Web search engine.
Rose and Levinson~\cite{Rose04} describe a framework for understanding the underlying goals of user searches.
Baeza-Yates et al.~\cite{Baeza-Yates05} analyze query log data and show several models about how users search and how users use search engine results.
Chau et al.~\cite{Chau05} study the information needs and search behavior of the users of a Web site search engine and compare them with those of general-purpose search engine users.
Kammenhuber et al.~\cite{Kammenhuber06} use client-side logs to evaluate user behavior in what they call Web search click-streams, i.e., search-induced clicks on the answer page provided by the search engine and the subsequently visited hyperlinked pages.

Nevertheless, characterizing the workload imposed by queries on typical Web search engines is crucial not only for evaluating search \textit{efficiency} in terms of, for instance, query
 response times, but also for designing predictive models.
This is our focus in this paper.
Beitzel et al.~\cite{Beitzel04} analyze hourly variations in query traffic and remark that the number of queries issued is substantially lower during non-peak hours than peak hours.
In our work, in addition to confirming this result for two different real-world vertical search engines, we characterize four key aspects of the workload imposed on these search engines, namely query and term popularity, number of terms per query, query interarrival times and query service times, with the goal of driving the design and evaluation of a performance model and its applicability to capacity planning~(see Section~\ref{section:workload}).

\subsection{Performance Modeling}
\label{section:relatedwork:performance}

Although many performance models exist for capacity planning of different systems~cite{Menasce04}, the availability in the literature of performance models is rather limited for Web search engines in general and in particular for vertical search engines.
Cacheda et al.~\cite{Cacheda04} present a case study of different architectures for a distributed information retrieval system, in order to provide a guide to approximate the optimal architecture with a specific set of resources.
Using a simulator based on an analytical model for query processing (similar to the one described in ~cite{Ribeiro-Neto98}), they analyze the effectiveness of a distributed, replicated, and clustered architecture simulating a variable number of workstations.
Nevertheless, their analytical model makes the simplifying assumption that service times are balanced if index servers manage a similar amount of data when processing a query.
Chowdhury and Pass~\cite{Chowdhury03} introduce a framework based on queueing theory for analyzing and comparing architectures for search systems in terms of their operational requirements: throughput, response time, and utilization.
Like in~cite{Cacheda04}, the proposed queueing model is not driven by observations from real workloads,  also assuming a perfect balance among the service times of index servers that process an equal number of documents per query.
Further, they do not verify the accuracy of their model by comparing its predictions with experimental results.

In contrast, we have previously found~cite{Badue07} that, even with a balanced distribution of the document collection among index servers, the heterogeneous use of disk cache in the homogeneous index servers  leads to imbalances in query service times (see Section~\ref{section:imbalance}).  Based on these previous findings, we here perform a detailed characterization of the workloads of two real-world vertical search engines, using their results to propose a performance model which, unlike
previous work, captures the impact of this imbalance on the average query system response time. 


\section{An Architecture for Vertical Search Engines}
\label{section:architecture}

\subsection{Cluster of Index Servers}
\label{section:cluster_of_index_servers}

\begin{figure}[htb!]
  \begin{center}
    \includegraphics[scale=0.75]{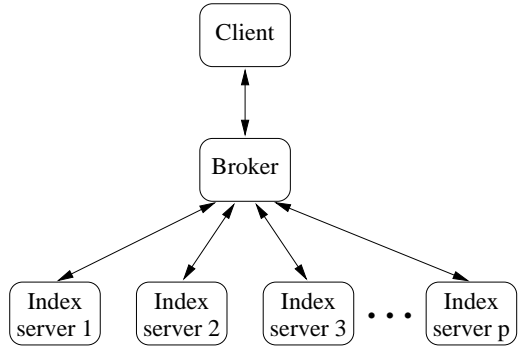}
    \caption{Architecture of a typical search engine.}
    \label{figure:architecture}
  \end{center}
\end{figure}

Modern search engines typically rely on computational clusters for query processing~\cite{Brewer01,Barroso03,Risvik03}.
Such clusters are mostly composed of a single broker and $p$~index servers.
Figure~\ref{figure:architecture} illustrates this architecture for a typical vertical search engine.

The broker receives user queries from client nodes and forwards them to the index servers, triggering the parallel query processing through the $p$~local subcollections.
Each index server searches its own local subcollection and produces a partial ranked answer.
These partial ranked answers are then sent to the broker where they are combined through an in-memory merging operation.
The final list of ranked documents is then sent back to the user.

We note that the described architecture is for a single query processing cluster, which constitutes the basic unit of modern search engines.
Large scale modern search engines basically replicate this cluster unit to support a higher query arrival rate~\cite{Brewer01,Barroso03,Risvik03}.

\subsection{Index Organization}

An inverted index is adopted as the indexing structure for each subcollection.
Inverted files are useful because they can be searched based mostly on the set
of distinct words in all documents of the collection.
They are simple data structures that perform well when the pattern to be
searched for is formed by conjunctions and disjunctions of
words~\cite{Ribeiro-Neto99,Witten99,Zobel06}.

The structure of our inverted indexes is composed of a {\em vocabulary} and a {\em set of inverted lists}.
The vocabulary is the set of all unique terms (words) in the document collection.
Each term in the vocabulary is associated with an inverted list that contains an entry for each document in which the term occurs.
Each entry comprises a document identifier and the within-document frequency $f_{t,d}$ representing the number of  occurrences of term $t$ within the document $d$.

The documents of the whole collection are uniformly distributed among the index servers.
We assign each document to an index server randomly, a policy that works well in balancing storage space utilization among servers~\cite{Badue05,Badue07}.
Let $n$ be the size of the whole document collection. A uniform distribution of the documents  among the $p$~index servers leads to  a size~$b$ of any local subcollection equal to $b=n
/p$. That is, the size of any local inverted file is $O(n/p)$.
This type of index organization, referred to as a {\em document partitioning}, is currently the de facto standard in all major search engines~\cite{Barroso03,Risvik03}.

\subsection{Parallel Query Processing}
\label{section:architecture:queryprocessing}

In this paper, we use the standard vector space model~\cite{Salton83} to rank the documents in the index servers.
Modern search engines also adopt link-based ranking, such as PageRank~\cite{Brin98}, combined with a complex text-based scoring function, to rank documents.
However, since link information is pre-computed offline as a global measure, its usage has only limited impact on performance.
Further, while we adopt here a simple ranking function, our model can capture the behavior of much more complex ranking functions, 
once key parameters are measured empirically.

In the vector model, queries and documents are represented as weighted vectors in a $t$-dimensional space, where $t$ is the  number of terms in the vocabulary of the collection.
Each term-document pair is weighted by the frequency $tf_{i,j}$ of term $k_i$ in document $d_j$ and the inverse document frequency $idf_{i}$ of the term $k_i$ among the documents in the whole collection.
The rank of a document with regard to a user query is computed as the cosine of the angle between the query and document vectors.
Using the {\em idf} weight implies that global knowledge about the whole collection is available at the index servers.
This can be accomplished if index servers exchange their local {\em idf} factors after the local index generation phase.
Each index server may then derive the global {\em idf} factor from the set of local {\em idf} factors~\cite{Ribeiro-Neto98}.

In our experiments, a client machine submits queries to the broker according to a query arrival distribution.
This broker then broadcasts each query to all index servers.
Once each index server receives a query, it retrieves the full inverted lists relative to the query terms, intersects these lists to produce the set of documents that contains all query terms (i.e., the conjunction of the query terms\footnote{Taking the conjunction of the query terms is now
standard practice on modern search engines.}), computes a relevance score for each document, sorts them by decreasing score, and
sends its ranked answer to the broker.
Each query term~$k_i$ is processed by decreasing $idf_i$, i.e., by increasing order of the number~$n_i$ of documents 
containing the term~$k_i$, thus leading to a significantly more efficient conjunction of their inverted lists.
As soon as the ranking is computed, the top ranked documents at each index server are transferred to the broker machine.
The broker is then responsible for combining the partial ranked answers received from the index servers through an in-memory merging operation.
The final list of top ranked documents is then sent back to the client machine.

We note that search engines that deal with large document collections perform a partial evaluation of the inverted lists relative to query terms instead of a full one~\cite{Ntoulas07}.
Nevertheless, in our experimental setup (see Sections \ref{section:capacityplanning} and \ref{section:applicability}), we do not introduce any bound on the number of entries of the inverted lists. By doing so, our performance model, parameterized from experimental measures, provides  conservative estimates of system performance, which are more adequate for capacity planning purposes. Morever, for  simplicity, we adopt a single processing thread at each index server.   
The use of multiple processing threads at index servers is left for future work.


\subsection{Imbalance in Query Service Times among Homogeneous Index Servers}
\label{section:imbalance}

In the architecture for parallel query processing, characterized by a local
partitioning of the document collection, the response time of a query is determined by the
service time of the slowest index server. As a consequence, any imbalance in service
times among index servers increases the response time of a query executed by the
cluster of servers. Therefore, it is critically important to avoid imbalance among
index servers if higher performance is to be achieved.

A common counter-measure against imbalance is to distribute the whole
collection of documents among homogeneous index servers (i.e., servers with identical hardware and software configuration) in a balanced way, such that
each server handles a similar amount of data for processing any given query.
At first glance, as a consequence of having similar data volumes handled at each server
for a given query, one would expect that service times at the homogeneous index
servers would also be approximately balanced. Indeed, this idealized scenario of
balanced service times is a usual assumption taken by the few previous theoretical models for Web
search engines available in the literature~\cite{Cacheda04,Chowdhury03}.

However,  we have previously shown that, in spite of the homogenous configuration and balanced amount of processing data, a non-negligible imbalance in per-query service time may arise within a cluster of homogeneous index servers~\cite{Badue07}. The key reason for this imbalance, which can be quite significant in certain scenarios, is the heterogeneous use of disk cache at the different index servers.  In other words, for a given query $q$,  even though all participating index servers tend to retrieve a similar amount of data, some index servers may experience much shorter service times for $q$ if the needed documents are found in the disk cache maintained in {\it  main memory} by the operating system. In contrast, the  remaining index servers, having to retrieve the documents from the hard disk, will experience much longer transfer delays, thus leading to an overall imbalance in service times.

In~\cite{Badue07}, we show that the heterogeneous disk cache behavior is directly affected by correlations between two key workload aspects, namely, non-uniform term popularity distribution and sizes of the corresponsing inverted lists. We also show that the main memory size at the homogeneous index servers, which defines the availability of resources for their disk caches, and the number of index servers are two other factors that impact disk cache behavior, and thus indirectly lead to per-query service time imbalance. In particular, the smaller the amount of main memory, the greater the chance that, while processing a query,
 some index servers might access their disk caches while the others have to access the hard disk.  Moreover, for a fixed amount of main memory, the larger the number of participating index servers, the higher the chance  that some of them will find documents in the disk cache. In both cases, the imbalance in service times increases.

\section{Workload Characterization}
\label{section:workload}

This section presents the first step of our capacity planning methodology, namely, the characterization of real search engine workloads.  The primary goal of this characterization is to build knowledge about key properties of real workloads 
and to draw insights into the design of cost-effective performance models. By cost-effective, we mean a model that is simple and yet accurate enough for practical capacity planning decisions in real setups. Additionally, we also provide models that best describe different workload characteristics that can be used in the design of realistic workload generators,  although addressing this particular research venue is outside the scope of the current paper.




Our characterization relies on past access logs containing queries to two vertical search engines no longer operational, namely TodoBR\footnote{TodoBr ({\texttt www.todobr.com.br}) is a trademark of
Akwan Information Technologies, which was acquired by Google Inc. in July 2005.} and Radix\footnote{Radix ({\texttt www.radix.com.br}) stopped its operation in 2001.}, both focused on the Brazilian Web. Additionally, we also have access to the workload of two much larger
Web search engines no longer operational, namely AllTheWeb\footnote{AllTheWeb (\texttt{www.alltheweb.com}) is a trademark of Fast Search \& Transfer company, which was acquired by Overture Inc. in February 2003.
In March 2004 Overture itself was taken over by Yahoo!} and AltaVista\footnote{AltaVista (\texttt{www.altavista.com}) was bought in 2003 by Overture Inc., which was acquired by Yahoo! in the same period.}, both focused on the whole Web.
We note that the current availability of query datasets from modern operational search engines is rather restricted because such data are usually considered sensitive by search engine operators. Thus, the access to {\it four} different query workloads, from four different real  search engines, sets our work in a priviledged position compared to most previous related efforts.

Table~\ref{tab:datasets} presents an overview of our datasets along several dimensions including total period covered by each log, total number of queries as well as numbers of unique queries and unique terms. The two general search engines present  much heavier query loads than
TodoBR and Radix, which is expected, since the former are worldwide search engines and the latter regional ones. In particular both AllTheWeb and Altavista have average daily loads that are two orders of magnitude heavier than the typical loads experienced by TodoBR and Radix.

\begin{table}[htb]
\caption{Overview of our  query datasets for vertical and general search engines.}
\label{tab:datasets}
\centering
\begin{tabular}{|l|c|c|c|c|}
\hline \hline
		& \multicolumn{2}{|c|}{Vertical} &  \multicolumn{2}{|c|}{General}\\ \cline{2-5}
                  & TodoBR & Radix & AllTheWeb  & Altavista\\
\hline \hline
Start Date & Jan 01 2003 & Jan 01 2003 & Sep 01 2003 & Sep 28 2001\\\hline
End Date   & Aug 31 2003 & Aug 31 2003 & Sep 21 2003 & Oct 03 2001\\\hline
Number of & \multirow{2}{*}{243} & \multirow{2}{*}{243} & \multirow{2}{*}{21} & \multirow{2}{*}{6}\\
days      &     &     &    &\\\hline
Number of & \multirow{2}{*}{6,806,795} & \multirow{2}{*}{19,934,196} & \multirow{2}{*}{25,080,586} & \multirow{2}{*}{7,169,365}\\
queries   &     &     &    &\\\hline
Number of        & \multirow{2}{*}{1,552,735} & \multirow{2}{*}{2,830,854} & \multirow{2}{*}{6,902,160} & \multirow{2}{*}{2,096,598}\\
unique queries   &     &     &    &\\\hline
Number of & \multirow{2}{*}{228,396} & \multirow{2}{*}{358,406} & \multirow{2}{*}{4,408,672} & \multirow{2}{*}{820,817} \\
unique terms   &     &     &    &\\\hline
Avg. number of  &  \multirow{2}{*}{28,012} & \multirow{2}{*}{82,034} & \multirow{2}{*}{1,194,314} & \multirow{2}{*}{1,194,893}\\
queries per day&     &     &    &\\
\hline \hline
\end{tabular}
\end{table}

The following subsections present our characterization of five relevant workload aspects, namely, query length, query popularity  and term popularity, query interarrival times and query service times. 
The design and validation of our performance model, driven by our characterization findings, is presented in Section~\ref{section:capacityplanning}.



\subsection{Query Collection}
\label{workload:querycollection}

We start by characterizing the query collection provided by each of the two vertical search engine datasets.
In particular, three important aspects that describe each collection are query length (i.e., number of terms per query), query popularity and term popularity.



The query length, i.e., the number of terms per query,
directly impacts the processing demand imposed for document retrieval at the index servers.  
We found a  median query length  equal to $2$ for both TodoBR and Radix datasets, whereas the average query length is $2.02$ and $1.91$, respectively.  These results suggest a significant trend towards queries composed of just one to a few terms. In fact, Table~\ref{tab:querylengths}, which shows the query length distributions,  confirms the prevalence of very short queries---i.e., queries containing at most two terms---for the two datasets.

\begin{table}[htb!]
\caption{
Distribution of query lengths in the vertical search engine datasets.}
\label{tab:querylengths}
\begin{center}
\begin{tabular}{|c|c|c|c|c|}
\hline \hline
Number of terms & TodoBR  & Radix \\ 
\hline \hline
1 &  0.32 & 0.35 \\ \hline 
2 & 0.41  & 0.43 \\ \hline 
$\geq$ 3 & 0.27  & 0.22 \\ \hline 
\hline \hline
\end{tabular}
\end{center}
\end{table}



\begin{figure}[htb!]
	\centering
	\subfigure[Query Popularity]
		{\includegraphics[scale=0.75]{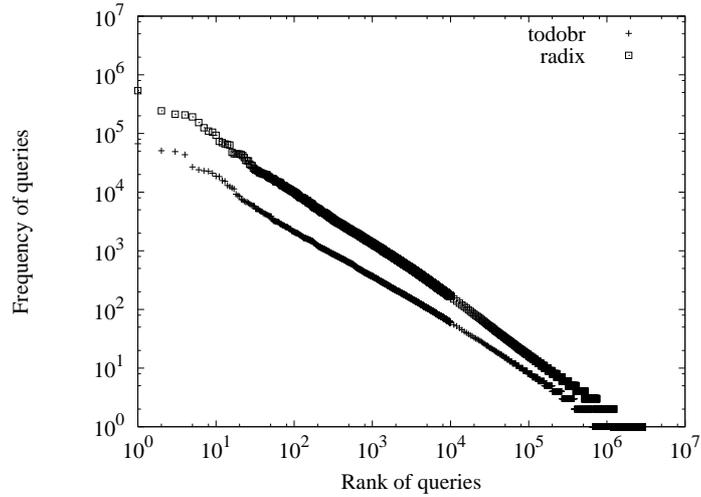}\label{figure:queryfreq}}\\
	\subfigure[Term Popularity]
		{\includegraphics[scale=0.75]{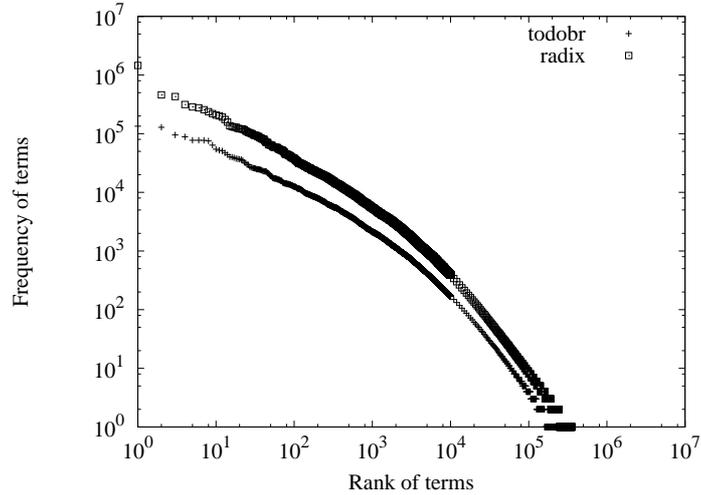}\label{figure:termfreq}}
	\caption{Distributions of frequency of unique queries and frequency of unique terms in the vertical search engine datasets, plotted in log-log scale.}
	\label{fig:freq_todobr_radix}
\end{figure}

 Next, we analyze the diversity of queries and of terms in queries.  Figures~\ref{fig:freq_todobr_radix}(a) and ~\ref{fig:freq_todobr_radix}(b) show the distributions of popularity (i.e., frequency) of unique queries and of unique terms in the TodoBR and Radix datasets, respectively.  Although the query datasets cover different query loads, the distributions of popularity of queries and of terms in queries are quite similar. All of them present strong skews towards the most popular elements, while still exhibiting  heavy tails. In fact, all of them follow Zipf distributions, as in Baeza-
Yates et al.~\cite{Baeza-Yates04,Baeza-YatesECIR05}. That is, given that $Prob(E_n)$ is the probability that the $n^{th}$ most popular element (query or term) occurs in the dataset, then  $Prob(E_n) \propto n^{-\alpha}$, where $\alpha$ is an empirical parameter.
By fitting a straight line to each curve in the log-log plots presented in Figures~\ref{fig:freq_todobr_radix}(a) and~\ref{fig:freq_todobr_radix}(b), we estimate the value of the parameter~$\alpha$ to be equal to $0.82$ and  $0.89$ for the distributions of query popularity in the TodoBR and Radix datasets, respectively. Similarly, we found values of $\alpha$ equal to $0.98$ (TodoBR) and $1.09$ (Radix) 
for the distributions of term popularity.
To illustrate the skew in these distributions, we verified that 1\% of the queries accounts for 41\% and  59\% of the requests in the TodoBR and Radix datasets, respectively. The skews observed  in the term popularity distributions are even stronger.

In practice, these skewed distributions of query popularity favor the use of disk caching at the index servers. The prevalence of a few terms per query combined with the strong skew in the term popularity distributions have similar implication. Ultimately, the heterogeneous use of disk cache at the index servers, exacerbated by a large number of servers or by a small amount of main memory,  leads to significant service time imbalances, as discussed in Section \ref{section:imbalance}. These imbalances must be taken into account when estimating
the performance of a search engine, at the cost of greatly underestimating its capacity needs, as previous work does~(see Section~\ref{section:relatedwork:performance}).  Our strategy to capture service time imbalance in a performance model is described in Section~\ref{section:capacityplanning:system}.

\subsection{Query Interarrival Times}
\label{section:workload:interarrivaltimes}

One key performance characteristic of the workload
 is the query arrival process, characterized by the distribution of query interarrival times. This distribution must be characterized in periods during which the arrival process is stable (e.g., fixed arrival rate). Otherwise, the aggregation of multiple workload behaviors might lead to unrepresentative distributions. Thus, our  first step towards characterizing interarrival times consists of analyzing the temporal evolution of query arrivals.

\begin{figure}[htb!]
\centering
	\includegraphics[scale=0.75]{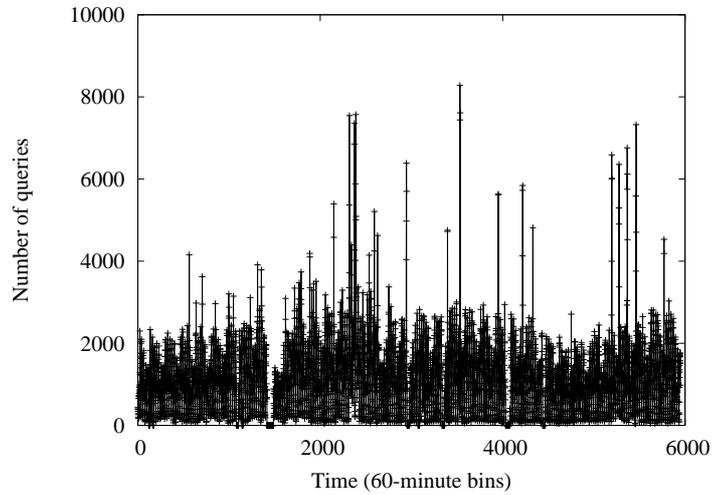}
	\caption{Query workload variation in the TodoBR dataset.}
	\label{fig:60minutes}
\end{figure}

\begin{figure}[htb!]
	\centering
	\subfigure[TodoBR]{\includegraphics[scale=0.75]{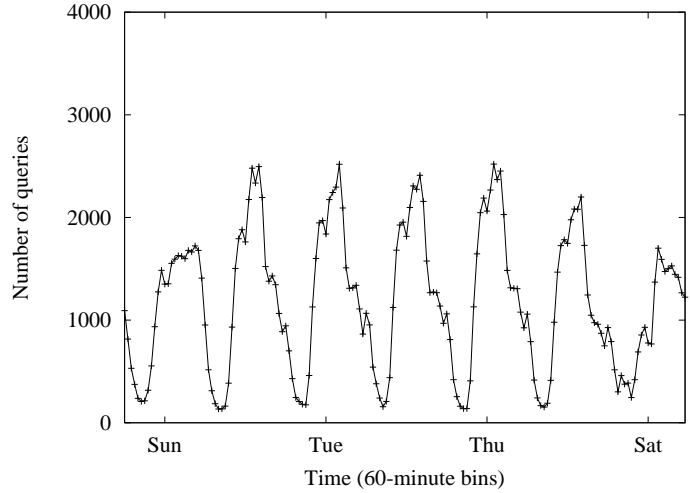}\label{sf:todobr}}
	\subfigure[Radix]{\includegraphics[scale=0.75]{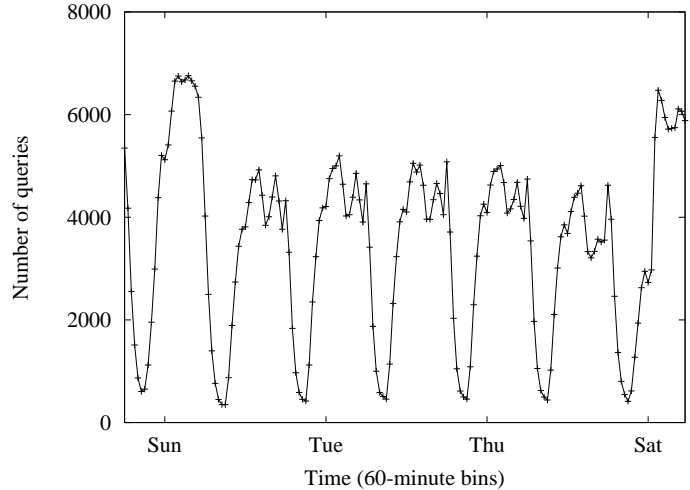}}
	\caption{Average number of queries over time (modulo one week) in the vertical search engine datasets.}
	\label{fig:1week}
\end{figure}

A periodic behavior in the number of query arrivals is observed in the two datasets, in different time scales. To illustrate this behavior, Figure~\ref{fig:60minutes} presents the number of query arrivals within 60-minute bins over the whole duration of the TodoBR dataset. The curve for Radix shows similar qualitative profiles, being thus omitted. 
Figure~\ref{fig:1week} shows the number of query arrivals per 60-minute bin, for each day of the week, averaged over all weeks, for the two datasets.
The query load in both datasets presents clear hourly load variations, with load peaks during the day and drops during the night. There are also significant load changes from working days and the weekend. Whereas in TodoBR, the load decreases during weekends, Radix experiences an opposite trend.
These results are summarized in Table~\ref{table:arrivalrate}, which  shows the average query arrival rate (in queries per second) over all periods covered by
each dataset. 

\begin{table}[htb!]
  \caption{ Average query arrival rate (queries/second) in the vertical search engine datasets.}
  \label{table:arrivalrate}
  \begin{center}
    \begin{tabular}{|c|c|c|c|}
      \hline\hline
      Day & \multicolumn{3}{|c|}{Dataset}\\ \cline{2-4}
      & TodoBR & Radix& Folded TodoBR \\
      \hline\hline
      Sunday & 0.48 & 1.88  & 16.27\\ \hline
      Monday & 0.69 & 1.36  & 23.58 \\ \hline
      Tuesday & 0.70 & 1.45  & 23.79  \\ \hline
      Wednesday & 0.67 & 1.40  & 22.77 \\ \hline
      Thursday & 0.70 & 1.40  & 23.80 \\ \hline
      Friday & 0.61 & 1.33  & 20.77  \\ \hline
      Saturday & 0.47 & 1.80  &  16.05 \\ \hline \hline
    \end{tabular}
  \end{center}
\end{table}


Before proceeding, we make an important observation that drives our next steps.
In spite of the number of available query datasets being rather limited, it is even harder to obtain datasets for the collection of documents used by Web search engines to generate an answer page for each received query.
This is because the set of collected documents is seen as strategic and proprietary information by the major Web search operators.
A similar trend characterizes collections of documents found in vertical search engines.
However, access to the document collection is  of fundamental importance to characterize query service times using real experiments, as we do in the next section, as well as to  the experimental validation of our performance model, presented in Section \ref{section:capacityplanning:validation}.

\vspace{\baselineskip}

Although we have access to four different query datasets (including those from two general search engines), we only have access to the document collection of one of them, namely the vertical search engine TodoBR. 
Restricting our experiments to the TodoBR dataset only would be constraining since, as shown in Figure \ref{fig:1week} and Table \ref{table:arrivalrate}, this search engine is somewhat light-loaded in terms of query arrival rates as compared to  Radix. 
We note, however, that, though relatively light-loaded, the TodoBR dataset has characteristics (distribution of the number of terms per query, skew in query popularity and term popularity, and daily/hourly load variations) that are qualitatively similar to those found in the other datasets, including those of general search engines, 
being thus representative of them in terms of these workload aspects.

\begin{figure}[htb!]
\begin{center}
\includegraphics[scale=0.75]{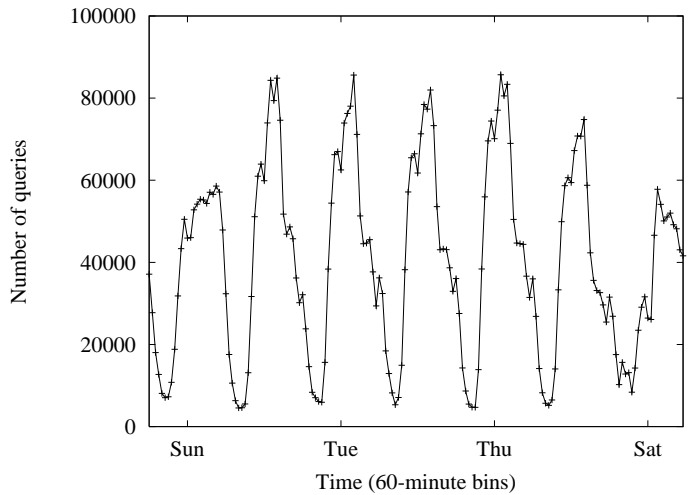}
\caption{Number of queries over time in the folded TodoBR dataset.}
\label{figure:loadvariationfoldedtodobr}
\end{center}
\end{figure}

Thus, in other to obtain a complete (query and document) dataset that is representative in terms of the key performance characteristics of other real datasets but has a heavier load, we apply a folding procedure to the TodoBR dataset. This procedure boosts the query arrival rate while still preserving the distribution models that best represent the performance characteristics under analysis. It consists of breaking the datasets into a number of consecutive equal-size time windows (e.g., 1 day, 1 week, 4 weeks, etc) and merging all queries falling into corresponding periods of all time windows. As an example, one could fold the dataset by considering a time window of 1 week and  merging all queries that arrive on each day of the week (e.g., Monday), using data from the entire dataset. 
The shorter the time window considered, the larger the load boosting factor.  
In this paper, we fold the original TodoBR dataset by considering a time window equal to 1 week, although other windows could have been used leading to similar (qualitative) results. Figure~\ref{figure:loadvariationfoldedtodobr} presents the daily load variation in the resulting folded TodoBR dataset, whereas the far right column in Table~\ref{table:arrivalrate} shows the average query arrival rates. The folded TodoBR dataset will be used, together with the corresponding document 
 collection, in the experiments discussed in Sections \ref{section:workload:servicetimes} and \ref{section:capacityplanning:validation}.

We finally turn to the characterization of the query interarrival times.
Given the high hourly and daily load variations observed in the original TodoBR and Radix datasets  as well as in the folded TodoBR dataset, we
characterize query interarrival times separately for a number of different 
heavy-loaded one-hour periods, during which arrival rates are stable. We evaluate the fittings provided by a set of well-known distribution models to the interarrival time distribution measured in each such period.  The distribution models considered are Exponential, Gamma, Weibull, Lognormal and Pareto. The fitting of each distribution is evaluated by computing the sum of the squares of the differences and the maximum difference (also known as Komogorov-Smirnov goodness of fit test~\cite{Chakravarti67}) between the measured interarrival time distribution and the best-fit provided by each of these well-known distributions. 

\begin{figure}[htb!]
\begin{center}
\includegraphics[scale=0.75]{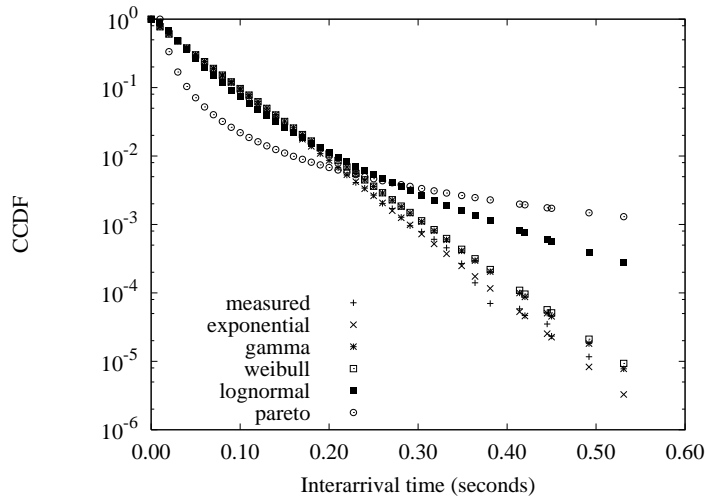}
\caption{Distribution of query interarrival times in the folded TodoBR dataset (one 1-hour period):
measured and predicted by various distribution models.}
\label{figure:fitinterarrivaltimes}
\end{center}
\end{figure}

For all time periods analyzed, in all three datasets, we found that the Exponential distribution\footnote{The probability density function (PDF) of the Exponential distribution is given by $f(x) = \frac{1}{\mu}e^{-\frac{1}{\mu}x}$, where $\mu$ is the distribution mean.} presents a fairly reasonable fitting compared to the Gamma and Weibull distributions, whereas the Lognormal and Pareto distributions fail to model the observed data accurately. In order to illustrate how close the fitting provided by the Exponential distribution is to the measured interarrival time distribution, in comparison with the alternative distributions, we show the curves of measured query interarrival times as well as the best-fits provided by each distribution in one selected high-load hour of the folded TodoBR dataset in Figure~\ref{figure:fitinterarrivaltimes}.

\subsection{Query Service Times} 
\label{section:workload:servicetimes}

To characterize query service times, we set up an experimental search engine, according to the architecture described in Section \ref{section:architecture},  in a dedicated environment consisting of a cluster of 8 homogeneous index servers and a single broker.  We used a test collection composed of roughly $10$~million Web pages collected by the TodoBR search engine from the Brazilian Web in~$2003$.  The collection was uniformly distributed over the $8$~index servers, resulting in a local subcollection of size $b = 1.25$~million  pages.  Each index server executes a simple ranking computation that combines text scoring with link analysis.  We also instrumented the index servers to collect and dump into a file the total time spent servicing each query.


We ran a series of experiments issuing queries extracted from the folded TodoBR dataset.  For each issued query, we collected the service time measured at each index server.
In particular, the following discussion refers to the service times collected for a set of $85,604$~queries, extracted from  a high-load one-hour period of that dataset, and measured  after warming up the index servers.

\begin{figure}[htb!]
\begin{center}
\includegraphics[scale=0.75]{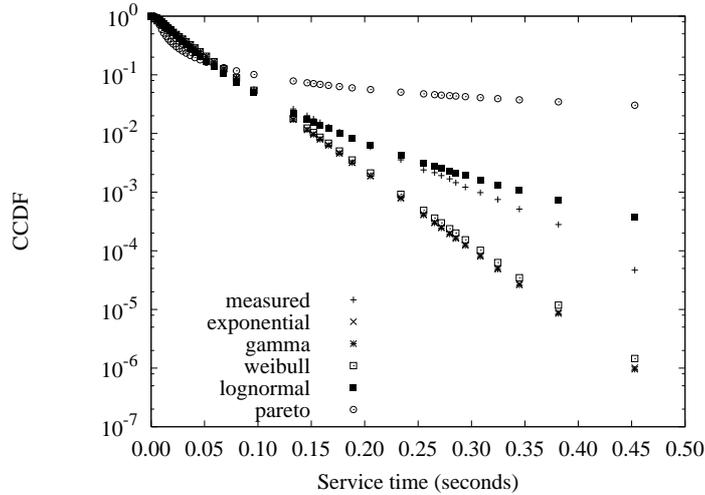}
\caption{Distribution of query service times at one index server of the (folded) TodoBR dataset: measured and predicted by various distribution models.}
\label{figure:fitservicetimes}
\end{center}
\end{figure}

We evaluate the fitting provided by well-known distribution models to the service time distribution measured in {\it each} index server.
As for the query interarrival times, we considered the fittings provided by Exponential, Gamma, Weibull, Lognormal and Pareto distributions, and evaluated them  using the same criteria discussed in Section \ref{section:workload:interarrivaltimes}. We found that Exponential, Gamma, Weibull and Lognormal distributions provided fairly reasonable fittings, whereas the Pareto distribution fails to model the observed data accurately. Figure~\ref{figure:fitservicetimes} illustrates the fittings of each distribution model to the service times measured at one index server.  Among the four best fittings, we choose to model query service times at each index server using an Exponential distribution.
This decision is based on two main reasons. Firstly,   our results demonstrate that the Exponential distribution approximates the observed  service times per index server within fairly reasonable bounds.  In fact, Figure~\ref{figure:fitservicetimes} shows that noticeable deviations emerge only for service times greater than 0.12 seconds, accounting for  a very small percentage (3\%) of the queries.  Secondly, the Exponential distribution is much simpler to model than the other ones, depending only on one parameter ($\mu$) given by the average service time.
Moreover, by comparing the fittings for all index servers, we found that the {\it same} Exponential distribution (i.e., same average service time and thus same distribution parameter $\mu$) fits reasonably the service times measured in {\it all} index servers.

Therefore, from the characterization of our two vertical search engine workloads, we learned that the skewed query and term popularity distributions combined with typically small query lengths may ultimately lead to significant service time imbalances among homogeneous index servers, and that query interarrival times and service times can be well approximated by Exponential distributions. Even more, query service times at different index servers follow the same Exponential distribution. These findings will drive our performance modeling strategy, which will be discussed in Section \ref{section:capacityplanning}.

\section{Performance Model for Vertical Search Engines}
\label{section:capacityplanning}

This section introduces the second step of our capacity planning methodology, namely, the performance model for vertical search engine architectures  (described in Section~\ref{section:architecture}).
Our goal is to have a model that can answer questions such as:
\begin{itemize}
\item[(i)] Given a collection composed of $n$~documents distributed over $p$~machines, what kind of average query response time guarantees can one  expect?
\item [(ii)] What kind of optimization in machine resources might yield a reduction in the average query response time to meet a service level objective defined by the manager of a vertical search engine?
\item[(iii)] What is the minimum number of replications of the cluster of index servers that will guarantee that, on average, the query response time in a peak period will not exceed the threshold defined by the manager of a vertical search engine?
\end{itemize}

Our capacity planning strategy relies on a queueing-based analytical model to estimate the {\it average query system response time}. Its design was driven by the empirical observation that a tool to be useful to operators of a vertical search engine should be easy to configure and to apply in practical scenarios. Thus, model simplicity is of the utmost importance, even if it comes at the cost of a {\it reasonable} compromise
in model accuracy. As will be shown, our model is simple, relies on easy-to-collect data, and still has reasonable accuracy in providing conservative estimates for 
capacity planning.

\subsection{Model Overview}
\label{section:capacityplanning:overview}

\begin{figure}[htb!]
\centering
\includegraphics[scale=0.75]{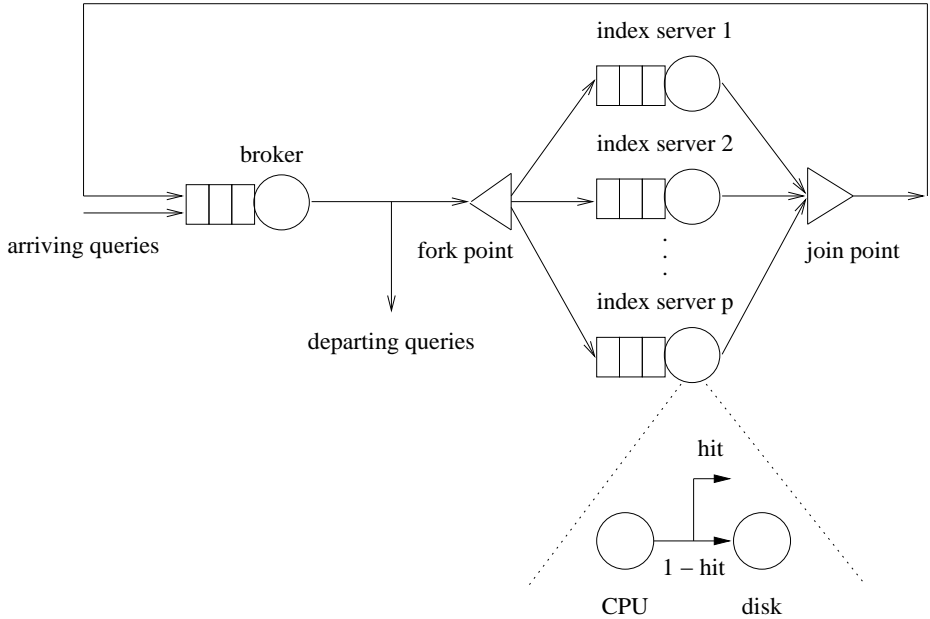}
\caption{Queueing network for a vertical search engine.}
\label{figure:qn}
\end{figure}

\vspace{\baselineskip}

The vertical search engine is represented by the queueing network described in Figure~\ref{figure:qn}. Before describing its main components, we introduce some basic definitions that are useful to understand:\\

\textbf{Definition 1:} The {\it service time} (or demand) of a query at a resource (e.g., a server) is defined as the time the query spends receiving service at the resource. \\

\textbf{Definition 2:} The {\it residence time} of a query at a resource  (e.g., a server) is defined as the sum of time the query spends waiting in the queue  to be served (i.e., {\it waiting time}) and its service time. \\

\textbf{Definition 3:} A {\it fork-join queueing network}~\cite{Menasce04} composed of a number of service centers (e.g., index servers) is used to model parallelism and concurrency in computer systems. When starting a concurrent processing stage, a task is split (i.e., {\it fork}) into a number of identical subtasks. Each subtask is then sent to a service center, where it is executed independently from the other subtasks. When a subtask finishes execution, it waits until all its sibling subtasks finish (i.e., {\it join}). Only at this moment, does the task complete execution and leave the network.  

\vspace{\baselineskip}

As represented in Figure \ref{figure:qn}, we model a vertical search engine as
an open queueing network composed of the broker and the subsystem of index servers.
We assume the index servers have homogeneous resources (as would be the case in several real scenarios) and that the collection of documents is uniformly distributed over all servers.
Based on the results presented in Section~\ref{section:workload}, we also assume that query service times at index servers are exponentially distributed.
Finally, the network connecting index servers and broker is typically a high-speed network and, as observed in our experiments, introduces negligible delays to the query system response time.
Therefore, it is not explicitly represented in our model.


In our search model, the index server subsystem is modeled as a fork-join queueing network composed of $p$ index servers.  The queueing discipline at each index server is FCFS (First-Come First-Served).
The behavior of this fork-join network, in which each task represents an arriving query and each service center models an index server, mimics the  parallelism in processing queries by the index servers and the synchronization introduced at the broker for combining partial results.
Mean Value Analysis (MVA)~\cite{Reiser80} offers an efficient solution for product-form queueing networks.
In particular, MVA can be used to produce performance estimates for each individual index server.
However, the fork-join feature violates the assumptions required by the exact MVA solution.
Thus, we use approximate MVA and bounding techniques~\cite{Menasce04} to solve the complete search model.

To process a query, an index server needs to retrieve the inverted lists related to the query terms from disk.
Thus, query service time at the index server is dominated by disk time and, possibly, CPU time.
However, due to the locality of reference of terms in the query log, an index server might find some or all of the inverted lists corresponding to the query terms in the disk cache (i.e., in main memory).
Thus,  some queries may not retrieve any data from disk at all.
In fact, during our validation experiments (see Section~\ref{section:capacityplanning:validation}), we found a non-negligible number of such queries in 
our workloads.


To capture the impact of disk cache, we refine our index server model as follows.
We separately model the average demands for CPU and disk (average CPU and disk times, respectively), as well as the probability of full disk cache hit (i.e., {\it all} inverted lists 
for a query are found in the disk cache) at an index server.
Note that the impact of partial disk cache hits is indirectly captured by the CPU and disk times. 
We also assume that queries may have different CPU times depending on whether they retrieve any data from the disk.
Given that queries are processed sequentially by each index server (see Section~\ref{section:architecture}), there is no queueing at any resource (CPU nor disk) of an index server.

The query residence time at the broker, which depends on the number of servers, consists of local processing for broadcasting the query to all index servers, receiving partial results from all servers, and merging the received partial results.
This residence time at the broker is relatively low compared to the average query system response time,
due to two basic reasons.
First, broker's operation is fully carried out using main memory, thus demanding only CPU time as opposed to an index server's operation that is composed of CPU and disk demands.
Second, all the tasks the broker executes are relatively simple tasks that do not take much CPU time.
It should be noted that the broker does not have to make ranking computations and does not have to execute algebraic operations, other than comparing document ranks.

Table~\ref{table:parameters} presents the system and workload input parameters as well as the output parameters of our model.

\begin{table}[htb!]
\caption{Input and output parameters of our model.}
\label{table:parameters}
\begin{center}
\begin{tabular}{|l|p{10cm}|}\hline\hline
Inputs  &  Description \\ \hline \hline
$p$ & Number of index servers \\ \hline
$\lambda$ & Query arrival rate \\ \hline
$S_{broker}$ & Average service time of a query at the broker for a cluster with $p$~index servers\\ \hline
$S_{hit}$ & Average CPU time at an index server for a query that finds all inverted lists for the query terms in the disk cache\\ \hline
$S_{miss}$ & Average CPU time at an index server for a query that retrieves data from disk\\ \hline
$S_{disk}$ & Average disk time of a query at an index server \\ \hline
$hit$ & Probability of a query finding all inverted lists for the query terms in  the disk cache  \\ \hline \hline
\multicolumn{2}{}{}\\
\hline\hline
 Outputs & Description \\ \hline \hline
$R$ & Average system response time of a query \\ \hline
$R_{cluster}$ & Average residence time of a query at the index server subsystem \\ \hline
$R_{broker}$ & Average residence time of a query at the broker for a cluster with $p$~index servers\\ \hline
$R_{server}$ & Average residence time of a query at an index server \\ \hline
$S_{server}$ & Average service time of a query at an index server \\ \hline
$U_{server}$ & Total resource utilization of a query at an index server \\
\hline\hline
\end{tabular}
\end{center}
\end{table}

\subsection{Model Solution}
\label{section:capacityplanning:solution}

This section describes our solution to the model in Figure~\ref{figure:qn}, which we use to estimate the average query system response time of a vertical search engine. Our main design goals are simplicity and reasonable accuracy, while
providing a conservative estimate on performance.
Moreover, we are particularly interested in solutions that deliver a good tradeoff for heavy load scenarios, when the search engine is approaching saturation (i.e., server resource utilization close to $100\%$).

\subsubsection{Index Server Model}
\label{section:capacityplanning:indexserver}

This section derives the average query residence time at one index server. 
Since we assume no queueing at any resource of an index server, a conservative assumption, we start by introducing an abstraction for each server as a single service center.
As a result, the average service time of a query at one index server is given by
\begin{equation}
\label{equation:sserver}
S_{server} = hit S_{hit} + (1 - hit )(S_{miss}  + S_{disk} ).
\end{equation}




Applying Little's result to an MVA equation for open networks~\cite{Menasce04}, we estimate the average residence time of a query at an index server as:
\begin{equation}
  \label{equation:rserver}
  R_{server} = \frac{S_{server}}{1 - \lambda S_{server}} \cdot
\end{equation}

We can also estimate the aggregated  utilization of an index server resources by the query:

\begin{equation}
\label{equation:rhoserver}
U_{server} = \lambda S_{server}
\end{equation}


Similarly, the average query residence time at the broker 
can be also easily estimated using MVA:
\begin{equation}
  \label{equation:rbrokerp}
  R_{broker} = \frac{S_{broker}}{1 - \lambda S_{broker}} \cdot
\end{equation}

Finally, the average query  system response time $R$ is computed as the sum of the average query residence times at the broker  and at the index server subsystem, i.e.: 
\begin{equation}
  \label{equation:r}
  R_{broker} + R_{server} \leq R,
\end{equation}

\noindent
where the inequality reflects the fact that $R_{server}$ considers residence time in a single index server.
%
\subsubsection{System Model}
\label{section:capacityplanning:system}

Recall that the index server subsystem is modeled as a fork-join network.
There is no known closed-form solution for fork-join networks with more than 2 queues.
Hence, the performance metrics of such networks must be computed using approximation and bounding techniques.
An easy lower bound on the  average query residence time in the fork-join subsystem is obtained by ignoring the synchronization delays and considering the average query residence time in the fork-join subsystem equal to the average query residence time at an  index server (see Equation~\ref{equation:rserver}).
Indeed, this is the solution adopted by Chowdhury and Pass~\cite{Chowdhury03} to estimate the average query system response time of a search system that, similarly to our work, is modeled as an open queueing network composed of a broker and a subsystem of index servers.
However, as the number of index servers increases, we expect a significant deviation from this lower bound due to the synchronization overhead, a result of the imbalance among the index servers,
as discussed in Section~\ref{section:imbalance}.

A number of approximations for queueing models with fork-join synchronization,  with various degrees of complexity and accuracy, are available in the 
literature~\cite{Flatto84,Nelson88,Baccelli89,Makowski94,Thomasian94,Yeung96,Varki96,Varki99,Lui98,Chen01,Varki01}.
Nelson and Tantawi~\cite{Nelson88} propose a very simple upper bound on the average response time for fork-join queueing networks with exponential interarrival times and exponential service times (as verified in Sections \ref{section:workload:interarrivaltimes} and \ref{section:workload:servicetimes}, respectively), which depends only on the number of index servers~$p$, and on the average query residence time at {\it one} server.
Given the $p^{th}$~harmonic number $H_p = 1 + 1/2 + 1/3 + \dots + 1/p$, the upper bound on the average query residence time at the index server subsystem is given by:
\begin{equation}
\label{equation:rcluster}
R_{cluster}  \leq  H_p R_{server}.
\end{equation}

The harmonic number $H_p$ is greater than $1$ and provides an approximation to the fact that the
residence time in the cluster is determined by the slower server.
As will be shown in the following, we found this upper bound---although quite simple---yields reasonably accurate estimates of the average query residence time at the index server subsystem.
Combining equations~\ref{equation:rserver}, \ref{equation:rbrokerp} and~\ref{equation:rcluster}, we obtain the following bounds on the average query system response time for our search engine:
\begin{eqnarray}
\label{equation:rsystem}
R_{server} + R_{broker}  \leq R \leq H_p  R_{server} + R_{broker}.
\end{eqnarray}
\noindent
We note that for a large number $p$ of index servers, the $p^{th}$~harmonic number $H_p$ converges to $\ln(p) + O(1)$, indicating that there is a logarithmic factor between the upper and lower bounds.

\subsection {Model Validation}
\label{section:capacityplanning:validation}

A series of validation experiments were executed in a dedicated environment consisting of  a cluster of $8$ homogeneous index servers and a single broker, 
with loads
in the range of 10 to 50 queries per second.
Each index server  runs on a $2.4$~gigahertz Pentium~IV processor with  $256$~megabytes  of main memory and a $120$~gigabytes ATA IDE hard disk.
The broker is an ATHLON~XP with a $2.2$~gigahertz processor and $1$~gigabyte of main memory.
All of them run the Debian Linux operating system kernel version~$2.6$.
Index servers and broker are connected by a $100$~megabits/second  high-speed network.
We used a test collection composed of roughly $10$~million Web pages collected by the TodoBR search engine from the Brazilian Web in~$2003$.
The inverted index for the whole collection occupies roughly~$12$~gigabytes.
The collection was uniformly distributed over the $8$~index servers, resulting in a local subcollection of size $b = 1.25$~million  pages.
Each index server executes a simple ranking computation that combines text scoring with link analysis. More sophisticated ranking computations can also be represented, given that our model parameters do not depend on specifics of the ranking computation.
The query dataset used in our tests is composed of $85,604$~queries in a high-load hour 
of the folded TodoBR dataset.


The values of the  model input parameters 
were easily obtained by retrieving statistics collected by the Linux operating system and made available at the $/proc$ pseudo-filesystem, during the experiment.
We collected the value of each parameter for each query in our test dataset, by averaging the results for all queries at the end of the experiment.
CPU and disk times for a query are collected from the $/proc/stat$ pseudo-file.
To estimate the fraction   of queries that found all inverted lists for the query terms in the disk cache ($hit$), we monitored the total number of sectors successfully retrieved from disk by each query, 
available in the I/O statistic field of the $/proc/diskstats$ pseudo-file.
Table~\ref{table:systemparameters} presents the model input parameter values obtained in our experiments.
The CPU and disk times at the index servers and hit probabilities are averages for all index servers.

\begin{table}[htb!]
  \caption{Model input parameter values.}
  \label{table:systemparameters}
  \begin{center}
    \begin{tabular}{|l|c|}
      \hline \hline
      Parameter & Value\\ \hline
      $p$ & ($\leq$) 8 servers \\ \hline
      $b$ & 1.25 million pages \\ \hline
      $S_{broker}$, $p=2$  &  $0.33$ ms  \\ \hline
      $S_{broker}$, $p=4$  &  $0.39$ ms  \\\hline
      $S_{broker}$, $p=8$  &  $0.52$ ms \\ \hline
      $S_{hit}$  & $9.20$ ms\\ \hline
      $S_{miss}$ & $10.04$ ms\\ \hline
      $S_{disk}$        & $28.08$ ms\\ \hline
      $hit$ &   $0.17$ \\
      \hline\hline
    \end{tabular}
  \end{center}
\end{table}

\begin{figure}[htb!]
 \begin{center}
 \includegraphics[scale=0.8]{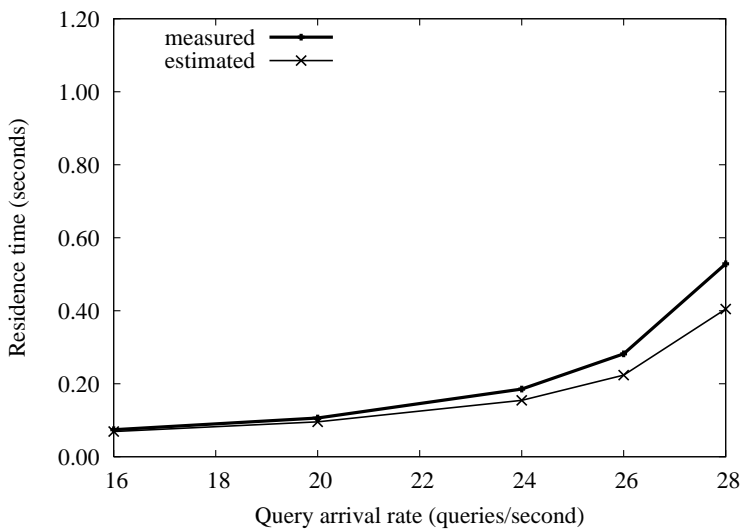}
 \caption{Average query residence time at an index server as a function of the query arrival rate ($p=8$).}
 \label{figure:rserver}
 \end{center}
 \end{figure}

Figure~\ref{figure:rserver} shows the average query residence time at an index server, averaged over all index servers, as a function of the total query arrival rate.
The ``estimated'' curve represents the results obtained with Equation~\ref{equation:rserver}, whereas the ``measured'' curve contains the average query residence times measured in all $8$ index servers.
As shown, our model captures reasonably well the average performance of a typical index server.
For an arrival rate of $28$~queries/second, the average utilization of the disk, the bottleneck resource at the index servers, is already almost~$79\%$, and the estimated aggregated utilization  of the index server resources ($U_{server}$) approaches~$92\%$.
Thus, the index server is approaching saturation.
For this load, the error introduced by  our model is only~$23\%$, reasonably small for response time estimates, as suggested in Menasce et al.~\cite{Menasce04}.
In the case of arrival rates higher than $28$~queries/second, our model predicts that the server saturates (i.e., server utilization is higher than $100\%$), which is indeed confirmed by our experimental measurements.

\begin{figure}[htb!]
  \begin{center}
    \includegraphics[scale=0.8]{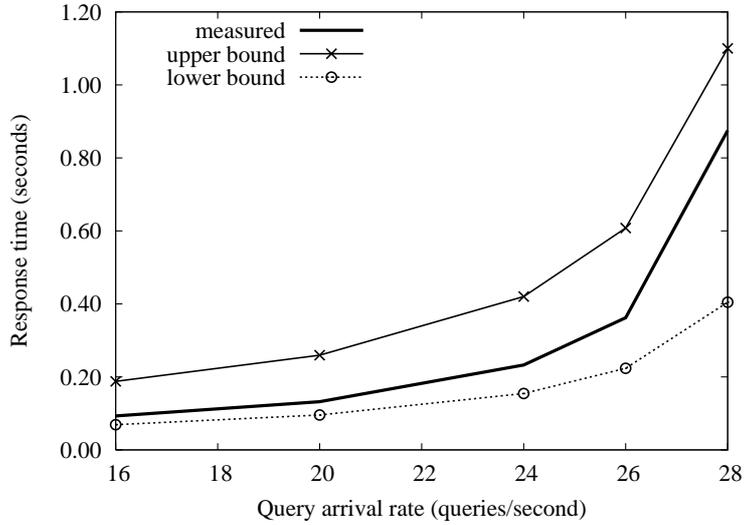}
    \caption{Average query system response time as a function of the query arrival rate ($p=8$).}
    \label{figure:rsystem}
  \end{center}
\end{figure}

\begin{figure}[htb!]
  \begin{center}
    \includegraphics[scale=0.8]{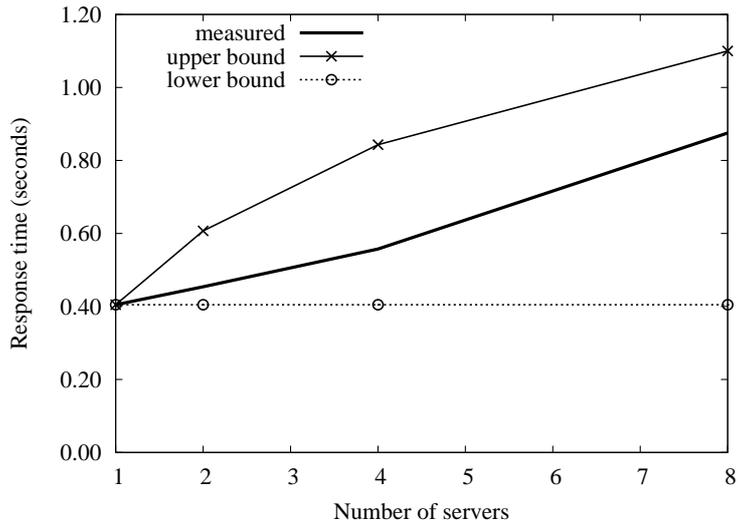}
    \caption{Average query system response time as a function of the number of index servers~$p$ ($\lambda = 28$~queries/second).}
    \label{figure:rsystemversusp}
  \end{center}
\end{figure}

We now turn to the validation of the average query system response time.
Figures~\ref{figure:rsystem} and~\ref{figure:rsystemversusp} show experimental results as well as the lower and upper bounds on the average query system response time, estimated via Equation~\ref{equation:rsystem}, as  a function of the arrival rate and of the number of index servers, respectively.
In Figure~\ref{figure:rsystemversusp}, in order to make a fair comparison, we kept the size of the subcollection~$b$ fixed (i.e., $1.25$~million pages per index server), varying only the total number of servers~$p$ and, indirectly, the size of the total collection~$n = pb$.
The upper bounds on the average query system response time for a number of servers~$p$ equal to $2$, $4$, and $8$ are $0.61$, $0.84$ and $1.10$, respectively. This confirms the logarithmic factor between the upper and lower bounds, as indicated at the end of Section~\ref{section:capacityplanning:system}.

As the results in this subsection show, the lower bound on the average  query system response time is a good approximation for systems with a small number of index servers and/or light loaded servers.
However, as either the load or the number of index servers increase, the measured average query system response time deviates significantly from the lower bound due to
the synchronization overhead.
This contrasts with previous work that disregards imbalance in query service times among homogeneous index servers~\cite{Chowdhury03,Cacheda04}.
In fact, we observe that the measured average query system response time approaches the upper bound  for larger number of index servers and heavier loads.
In particular, for $p=8$~index servers and an arrival rate $\lambda = 28$~queries/second (i.e.
close to saturation), the approximation error is only~$20\%$.
Therefore, the upper bound provides  a simple-to-compute and yet reasonably accurate  approximation of the average query system response time of
heavy-loaded (and thus perhaps more realistic) vertical search engines. The achieved accuracy should be enough for cost-effective capacity planning decisions on
such vertical search engines.

\section{Performance Model Applicability}
\label{section:applicability}

\vspace{\baselineskip}

In this section, we discuss how to use our model for the capacity planning of vertical search engines.
This is important because the number of index servers in a real cluster is usually large and maintaining large clusters operational is quite expensive.
Furthermore, our model is easy to use and depends only on information that is usually available in both search logs and standard operating systems, as described in Section~\ref{section:capacityplanning:overview}.

We consider ``what if'' questions, concerning future scenarios for a  vertical search engine.
We assume a collection of $1$~billion pages. 
We also assume that each index server stores a subcollection of size $b=10$~million pages---the largest collection we have available for experimentation, which requires $100$~index servers to host the whole collection.
To obtain the parameters of the index server model, we executed experiments in a reference system composed of a {\em single} index server running on a
machine with configuration equivalent to the index server described in Section~\ref{section:capacityplanning:validation}. 

To analyze the impact of adopting a larger main memory on the query system response time, we also executed experiments in this index server with
main memory four times larger than the reference machine.
Note that we needed a single (upgraded) index server to collect the new parameter values, but once
we have it we can predict performance with various numbers of servers.
We used a test collection composed of roughly $10$~million Web pages collected by the TodoBR search engine from the Brazilian Web in~$2003$, and a query dataset composed of $85,604$~queries in a high-load hour of the folded TodoBR dataset (see Section~\ref{section:workload}).

Based on these experiments, we determined parameter values for the index server model, such as the average CPU and disk times per query, and the probability of full disk cache hit at an index server for a query,  
while considering different main memory sizes.
To obtain the average query service time at the broker as a function of $100$~index servers ($S_{broker}$, $p=100$), we fit a straight line to the values of $S_{broker}$ ($p = 2, 4, 8$) estimated  during our validation experiments (see Table~\ref{table:systemparameters}).
We found an accurate fitting (coefficient of determination of $R^2 = 9.999870 \times 10^{-1}$) given by
$S_{broker} = 3.18 \times 10^{-2}  p +  0.265$ milliseconds, which we used to derive
$S_{broker} = 3.45$~milliseconds for $p=100$.
Table~\ref{table:newsystemparameters} presents the new parameter
values used in our example.

\begin{table}[htb!]
 \caption{New model input parameter values used in our example.}
 \label{table:newsystemparameters}
 \begin{center}
   \begin{tabular}{|l|c|c|c|c|}
     \hline \hline
     Parameter & \multicolumn{4}{|c|}{Value}\\ \cline{2-5}
     & Reference   & 2x the      & 3x the      & 4x the \\
     & size for    & reference   & reference   & reference \\
     & main memory & main memory & main memory & main memory\\\hline\hline
     $p$ & \multicolumn{4}{|c|}{$100$ servers} \\ \hline
     $b$ & \multicolumn{4}{|c|}{$10$ million pages} \\ \hline
$S_{broker}$ & \multicolumn{4}{|c|}{$3.45$ ms} \\ \hline
     $S_{hit}$ & $28.23$ ms & $33.38$ ms & $34.57$ ms
& $34.68$ ms  \\ \hline
     $S_{miss}$ & $35.31$ ms & $33.77$ ms & $32.66$ ms
& $32.04$ ms  \\ \hline
     $S_{disk}$ & $66.03$ ms & $35.89$ ms & $30.48$ ms &
$26.14$ ms  \\ \hline
     $hit$ & $0.02$ & $0.09$ & $0.15$ &  $0.18$  \\
     \hline\hline
   \end{tabular}
 \end{center}
\end{table}

Once the model parameters are computed, one can apply the model to derive the performance metrics of interest.\\[0.3cm]

\begin{figure}[htb!]
\centering
\includegraphics[scale=0.75]{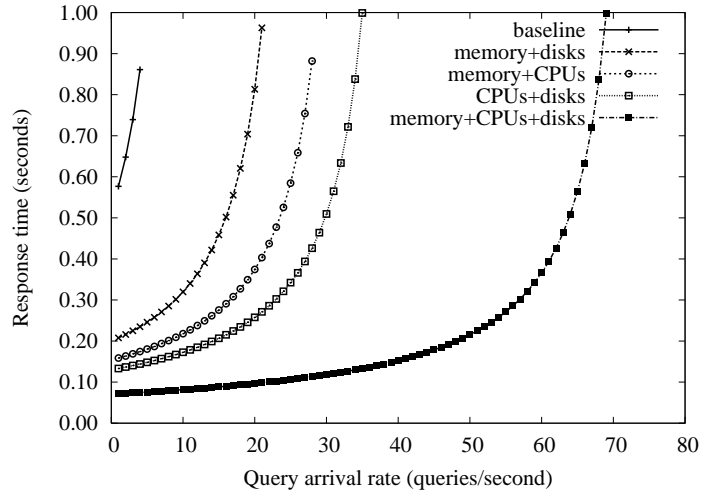}
\caption{Upper bound on the average query system response time as a function of the query arrival rate derived in our example.}
\label{figure:example}
\end{figure}

{\sf Case Study - The manager of a vertical search engine wants to guarantee that the average query system response time will not exceed 300 milliseconds. Further, by replicating clusters of servers the manager plans to support a combined arrival rate of $200$ queries per second.}\\[0.2cm]
To estimate the behavior of the parameters involved in this case study, we apply our capacity planning approach to the problem. Using the model with the parameters described in the second column of Table~\ref{table:newsystemparameters} (for an index server with the reference size for main memory), we calculate the upper-bound on the average query system response time as a function of the query arrival rate. The result, illustrated by the ``baseline'' curve in Figure~\ref{figure:example}, indicates that the baseline exceeds the threshold of $300$ milliseconds per query established by the manager, even at very low query arrival rates.

We want to evaluate what kind of optimization in the resources of index servers might yield a reduction in  the average query system response time to less than $300$~milliseconds.
For this, we consider five scenarios.
Also in Figure~\ref{figure:example}, the curves labeled  ``memory+disks'', ``memory+CPUs'', ``CPUs+disks'' and ``memory+CPUs+disks'' represent the upper bound on the average query system response time for four scenarios where:  (i) main memories are four times larger and disks are four times faster; (ii) main memories are four times larger and CPUs are four times faster; (iii) CPUs and disks are both four times faster; and (iv) main memories are four times larger, and CPUs and disks are both four times faster,  respectively.
Furthermore, we evaluate the impact of application-level caching on the response times predicted by our model based on recent experimental results found in the literature.
These are the five scenarios analyzed in the following.\\[0.3cm]
{\sf Scenario 1 - Main memories are four times larger and disks are four times faster.}\\[0.2cm]
In the first scenario, we want to evaluate the impact on query system response time of main memories that are four times larger and disks that are four times faster.
This is reflected in the  model parameter by dividing the disk time $S_{disk}$ (described in the fifth column of Table~\ref{table:newsystemparameters} for an index server with
main memory four times larger than the reference size) by a factor of four.
Solving the model with the new parameters yields the new average response time upper bound.
The results show that the upper bound on the average query system response time decreases significantly, with gains that reach approximately  $4$~times over the baseline system when it is approaching the point of saturation ($\lambda=4$~queries/second).
The reason for these gains is that the  probability of a query finding all inverted lists in  the disk cache increases with the size of the main memory, thus accelerating disk I/O.
Indeed, when the main memory increases by a factor of four in our experiments,  the probability of full disk cache hit increases by a factor of nine ($0.18 / 0.02$) and the demand for disks decreases by a factor of $2.53$ ($66.03 / 26.14$), as indicated in Table~\ref{table:newsystemparameters}.
Nevertheless, the upper bound still exceeds the defined threshold even at light loads.\\[0.2cm]
{\sf Scenario 2 -  Main memories are four times larger and CPUs are four times faster.}\\[0.2cm]
In the second scenario, we want to assess the impact on query system response time of main memories that are four times larger and CPUs that are four times faster.
The way of modeling the new CPUs is to divide the CPU times $S_{hit}$ and $S_{miss}$ (described in the fifth column of Table~\ref{table:newsystemparameters} for an index server with
main memory four times larger than the reference size)
by a factor of four.
Using the model, we calculate the new average response time upper bound.
The results indicate that the configuration with faster CPUs outperforms the configuration with faster disks, with gains that reach approximately $5$~times over the baseline system when the latter is is approaching saturation  ($\lambda=4$~queries/second).
There are two reasons for these gains.
First, the demands for CPUs are larger than the demands for disks in the configuration with main memory four times
larger than the reference size (see fifth column of Table~\ref{table:newsystemparameters}).
This implies that the optimization of CPUs yields a higher reduction in  the average query system response time than the optimization of disks.
Second, the optimization of CPUs improves both the performance of index servers and the performance of the broker, while the optimization of disks benefits only the index servers.
Nevertheless, the upper bound still exceeds the defined threshold even at somewhat light loads ($\lambda \geq 17$ queries/second).\\[0.2cm] 
{\sf Scenario 3 -  CPUs and disks are both four times faster.}\\[0.2cm]
In the third scenario, we want to verify the impact on query system response time of CPUs and disks that are both four times faster.
For modeling the new resources, we divide the CPU and disk times (described in the second column of Table~\ref{table:newsystemparameters})
by a factor of four.
The solution of the model with the new parameters yields the new average response time upper bound.
The results show that the configuration with faster CPUs and disks outperforms the configurations with larger main memories (in Scenarios~$1$ and $2$), with gains of approximately $6$~times  over the baseline system when it is approaching the saturation point ($\lambda=4$~queries/second).
Nevertheless, the upper bound still exceeds the defined threshold at moderate to high loads ($\lambda \geq 22$ queries/second).\\[0.2cm]
{\sf Scenario 4 -  Main memories are four times larger, and CPUs and disks are both four times faster.}\\[0.2cm]
In the fourth scenario, we want to evaluate the impact on query system response time of main memories that are four times larger, and CPUs and disks that are both four times faster.
This is reflected in the  model parameters by dividing the CPU and disk times
(described in the fifth column of Table~\ref{table:newsystemparameters} for an index server with
main memory four times larger than the reference size)
by a factor of four.
The results indicate that the upper bound on the average query system response time is equal to $286$~milliseconds at an arrival rate of $56$~queries/second,
which satisfies the service level objective for the search engine by bounding the average response time to less than $300$~milliseconds per query.
The results also show a remarkable reduction in  the upper bound on the average query system response time, with gains  that reach approximately $12$~times over the baseline system when the latter is approaching its saturation point ($\lambda~=~4$~queries/second).



\begin{figure}[htb!]
\centering
\subfigure[Reference main memory size]{\includegraphics[scale=0.47]{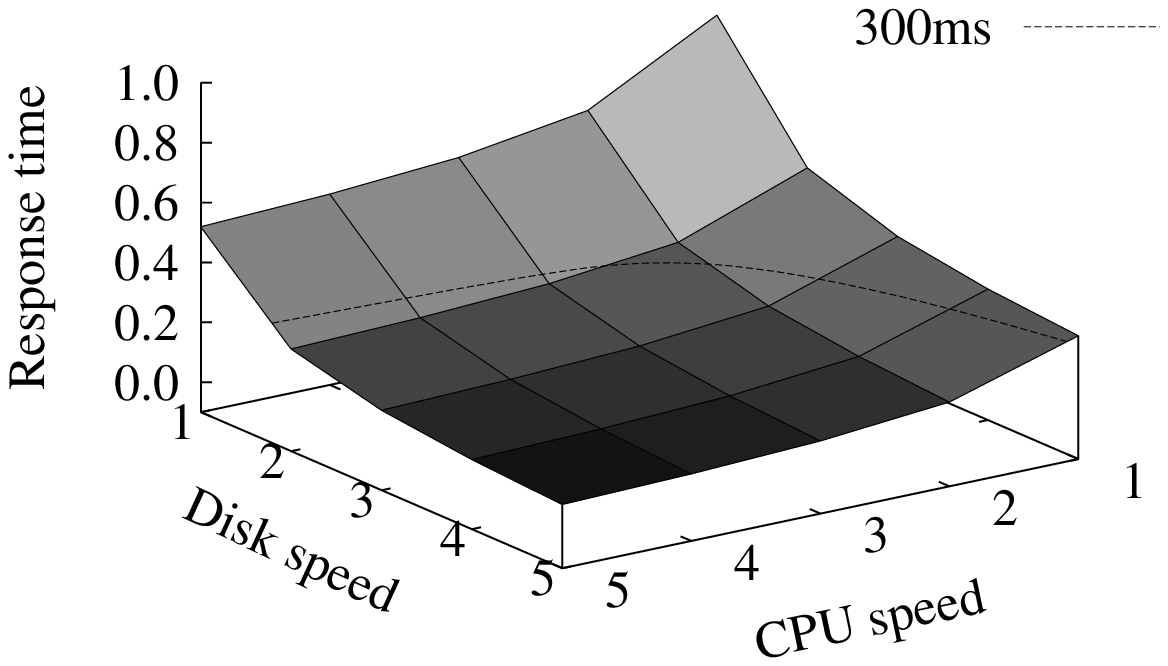}}
\subfigure[2x the reference main memory size]{\includegraphics[scale=0.47]{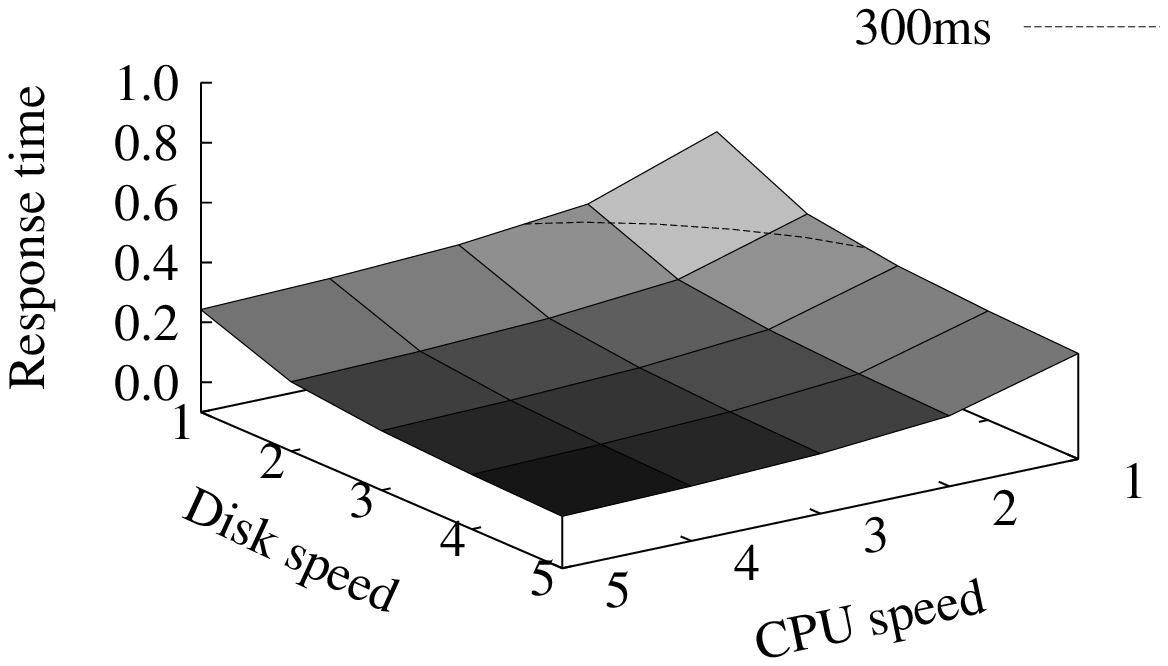}}
\subfigure[3x the reference main memory size]{\includegraphics[scale=0.47]{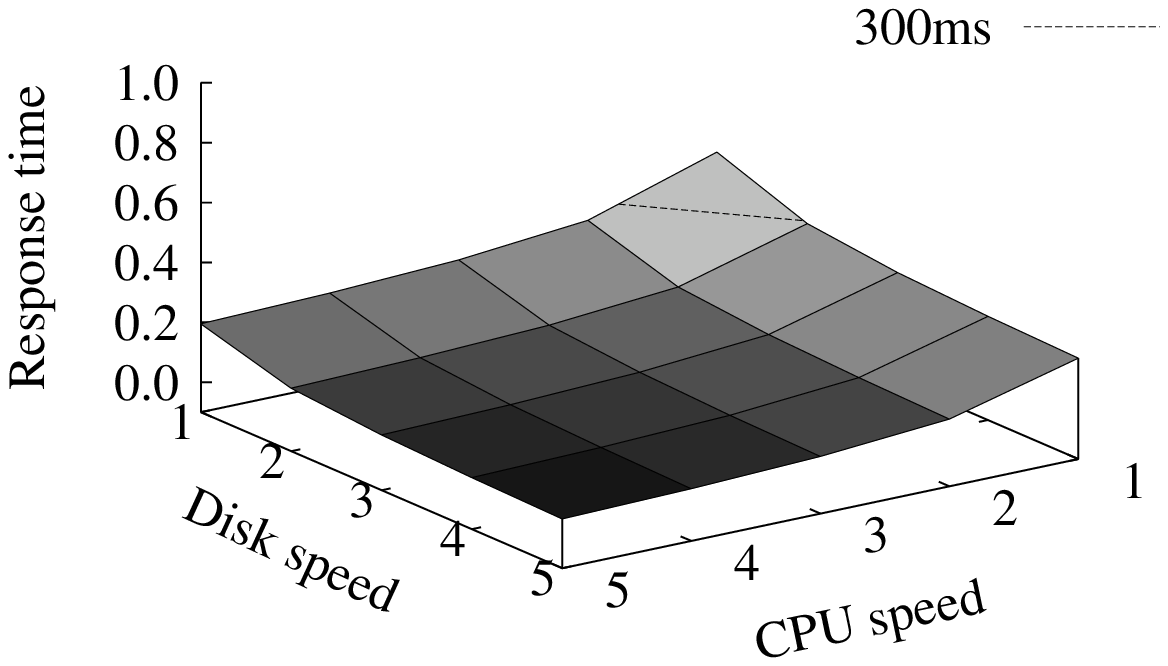}}
\subfigure[4x the reference main memory size]{\includegraphics[scale=0.47]{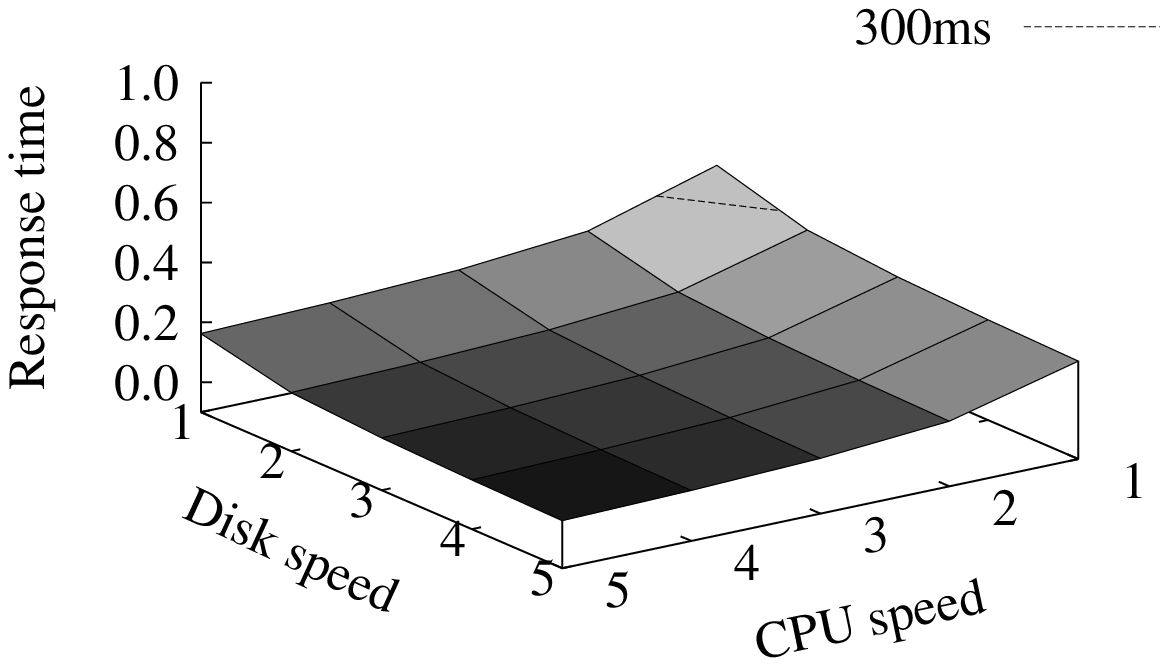}}
\caption{Upper bounds on the average query system response time (in seconds) as a function of main memory size, CPU speed (x times faster) and disk speed (y times faster),  for a fixed arrival rate ($4$ queries/second).}
\label{figure:example_3d}
\end{figure}

We still have to meet the goal of supporting an arrival rate of $200$~que-ries/second.
For supporting a higher query arrival rate, the cluster of index servers is usually replicated~\cite{Brewer01,Barroso03,Risvik03}.
Replication involves relatively small performance overheads, and approximately linear gains in the supported query arrival rate can be expected as a function of the number of mirrored systems.
The objective of supporting an arrival rate of $200$~queries/second can be achieved by creating $4$~replicas of the cluster of $100$~index servers, each replica supporting an arrival rate of $56$~queries/second and guaranteeing a query system response time of $286$~milliseconds.
Therefore, our model indicates that a cluster composed of  $400$~index servers ($4$~cluster replicas $\times$ $100$~index servers in a cluster) would achieve the desired performance of $200$~queries/second.
Some speculations suggest that  large scale search engines may indeed adopt clusters with several thousands of machines, but, to the best of our knowledge, there is no publicly available data to support this information.
If each index server handles a larger subcollection, the number of total servers in a cluster could be smaller.
Thus, one must be cautious before extrapolating our illustrative results to an arbitrary cluster of any search engine, in particular
large scale public Web search engines with a much larger scope than typical vertical search engines.
In this case, for other document collections, for other types of machines, or for other information retrieval algorithms, the parameters of our model could still be estimated experimentally and applied following the proposed methodology.\\[0.2cm]
{\sf Scenario 5 - Evaluating the impact of upgrade decisions.}\\[0.2cm]
%
We now use our model to analyze the impact on the overall performance of the system caused by alternative upgrade decisions, i.e. the influence on the overall performance of varying main memory sizes, CPU speeds, and disk speeds.
Figures~\ref{figure:example_3d}(a) to (d) show the upper bounds on the average query system response time as a function of main memory size, CPU speed and disk speed,  for a fixed arrival rate ($4$ queries/second).
The baselines for ``disk speed'' (i.e. disk speed equaling~1) and ``CPU speed'' (i.e. CPU speed
equaling~1) in these figures correspond to the disk and CPU speeds for which the model parameter values are shown in Table~\ref{table:newsystemparameters}. The results for different memory sizes are computed using the parameter values shown in columns $2-5$ in the same table.

We observe in Figure~\ref{figure:example_3d}(a) that response time decreases more rapidly with the increase of disk speed than with the increase of CPU speed.
This is a consequence of the much larger time for disk access to retrieve the inverted lists related to query terms, as compared to the case in which the lists are found in the disk cache. Since the available memory size is rather limited in this case, the demands for disk access are higher than demands for CPU processing, thus leading to increased response times.
Nevertheless, as the main memory size increases, as shown in  Figures~\ref{figure:example_3d}(b) to (d), the system faces a contrasting trend: average response time decreases with the increase of CPU speed, decreasing at much slower rate with disk speed.
When the main memory size is relatively large as compared to the size of the local index stored at the servers, there is more memory capacity available for the operating system to perform disk caching operations.
In this way, CPU demands become higher than disk demands contributing to the decrease of response time.
Note that results in this scenario consider only disk caching.\\[0.2cm]

\noindent
{\sf Scenario 6 -  Influence of application-level caching of query results at the broker.}\\[0.2cm]
Although our model does not incorporate application-level caching of query results, we built upon recent results from the work presented in~\cite{Baeza-Yates07}, in order to evaluate the influence on query system response time of caching query results at the broker.
Baeza-Yates et al.~\cite{Baeza-Yates07} study the tradeoffs in designing caching systems for Web search engines, including caching of answers to a particular query.
The strategy of caching query results is to keep in memory the list of documents associated with a given query.
This scheme allows the broker to answer recently issued queries at a very low cost, since it is not necessary to process these queries.

Actually, based on the Equation~\ref{equation:rsystem},
the upper bound on the average query system response time for our search system with caching of query results can be estimated as:
\begin{eqnarray}
\label{equation:rsystemcache}
R \leq \bigg(H_p  \frac{S_{server}}{1 - \lambda S_{server}} + R_{broker}\bigg) (1 - hit^{result}) + \frac{S_{broker}^{cache_{hit}}}{1 - \lambda S_{broker}^{cache_{hit}}} hit^{result}
\end{eqnarray}
where $S_{broker}^{cache_{hit}}$ is the average service time at the broker for a query that finds its results in the application-level cache and $hit^{result}$ is the probability of hit in the cache of query results at the broker.

\begin{figure}[htb!]
\centering
\subfigure[Reference main memory size]{\includegraphics[scale=0.47]{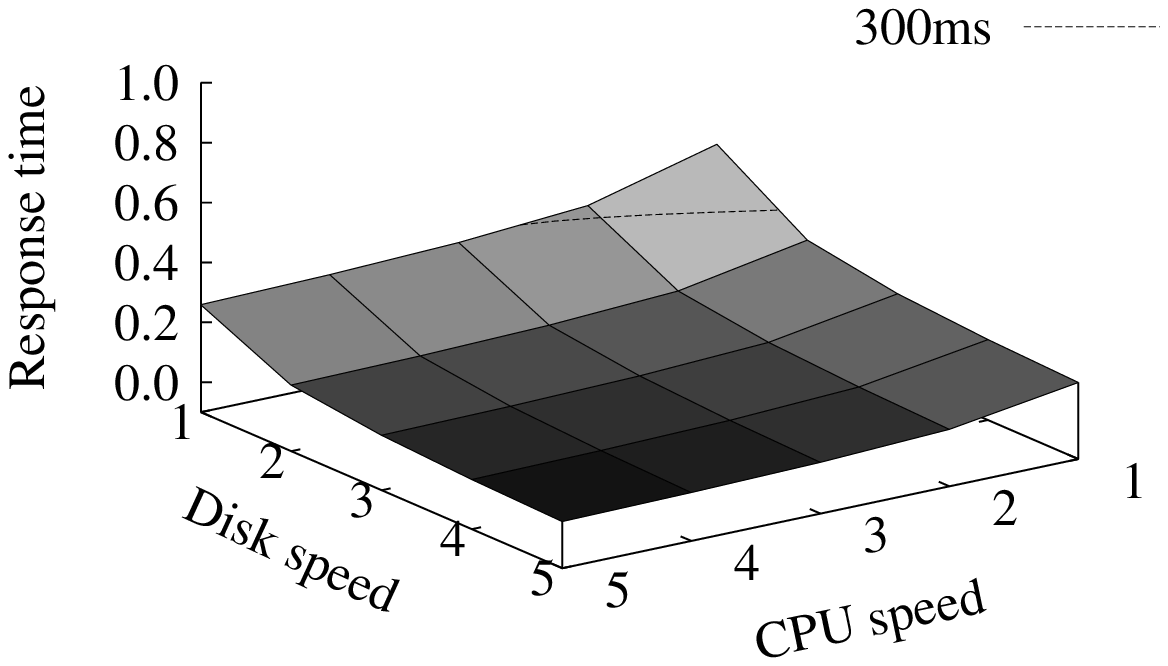}}
\subfigure[2x the reference main memory size]{\includegraphics[scale=0.47]{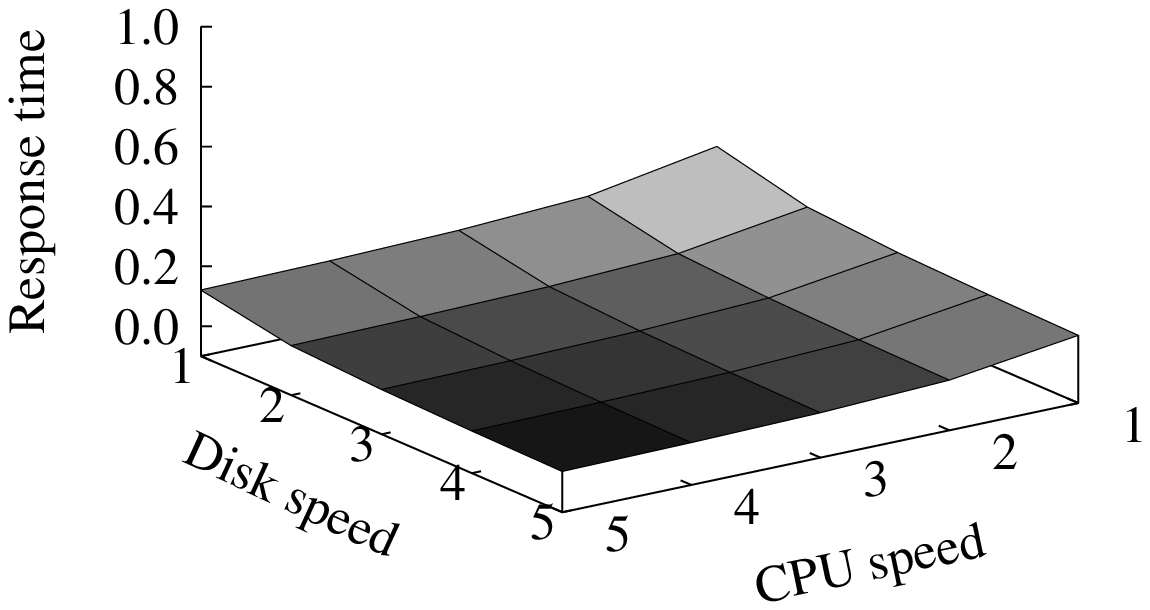}}
\subfigure[3x the reference main memory size]{\includegraphics[scale=0.47]{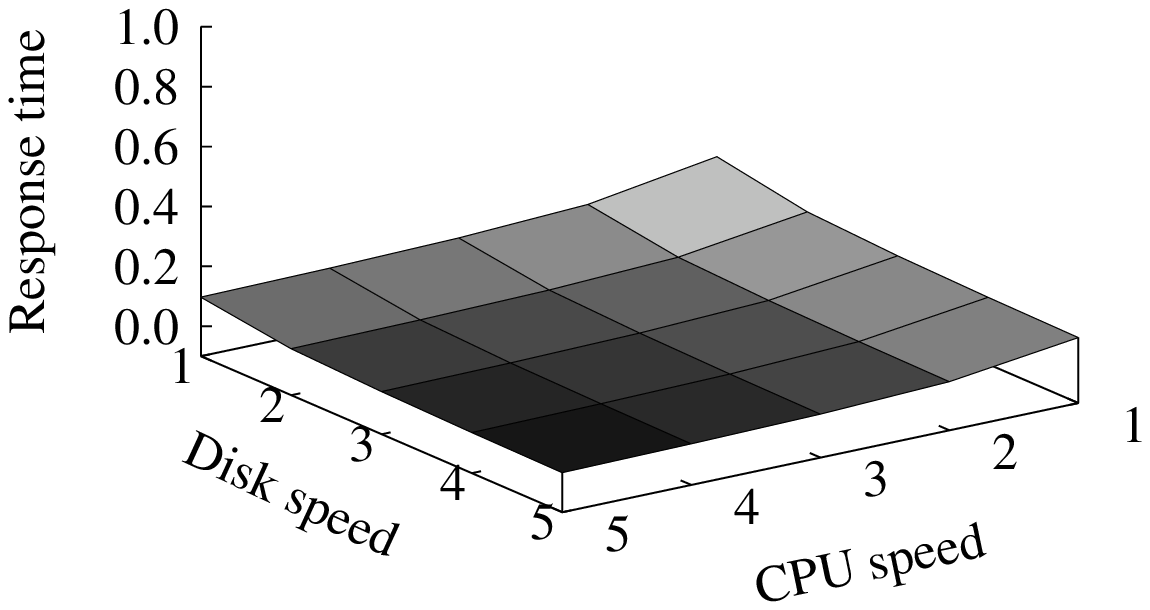}}
\subfigure[4x the reference main memory size]{\includegraphics[scale=0.47]{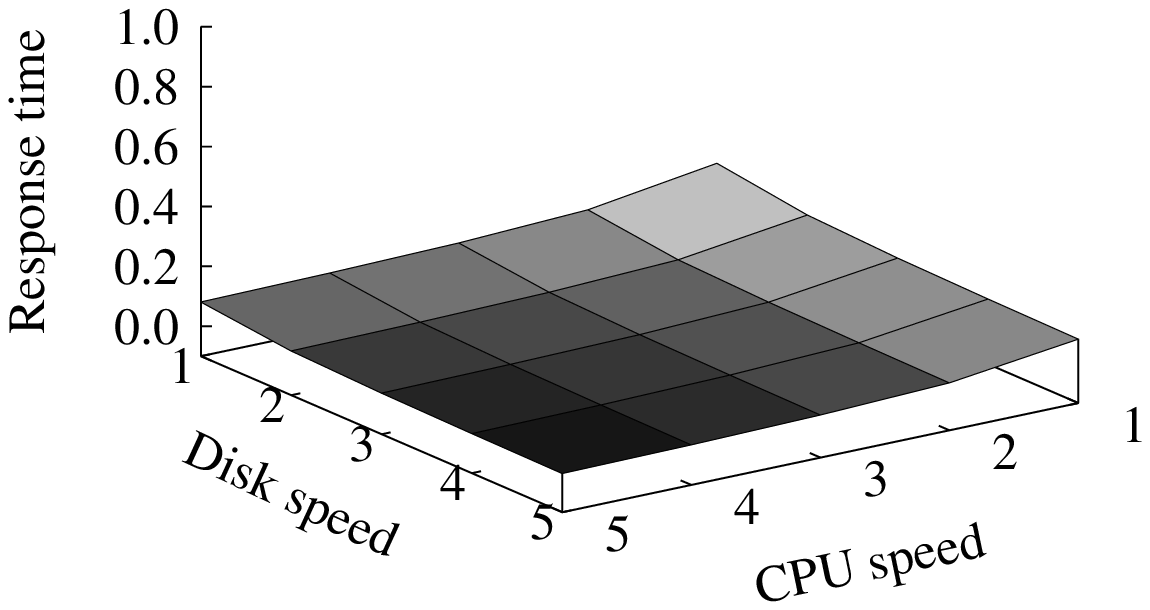}}
\caption{Upper bounds---that consider application-level cache of query results at the broker---on the average query system response time (in seconds) as a function of main memory size, CPU speed (x times faster) and disk speed (y times faster),  for a fixed arrival rate ($4$ queries/second).}
\label{figure:example_3d_cache}
\end{figure}

Using logs of one year of queries submitted to \url{http://www.yahoo.co.uk} from November $2005$ to November $2006$,
Baeza-Yates et al.~\cite{Baeza-Yates07} estimated experimentally a cache hit ratio of $0.50$, when considering a cache of query results with infinite memory, and an average time of $0.069$ milliseconds to return a stored answer for a query.
Applying these values to Equation~\ref{equation:rsystemcache} ($hit^{result} = 0.50$ and $S_{broker}^{cache_{hit}}=0.069$ milliseconds), we calculate the new upper bounds on the average query system response time, which now consider application-level caching of query results at the broker as a function of main memory size, CPU speed and disk speed, for a fixed arrival rate ($4$ queries/second), as shown in Figures~\ref{figure:example_3d_cache}(a) to (d).
We observe a significant decrease in response times through all system configurations as compared to the performance presented in  Figures~\ref{figure:example_3d}(a) to (d), in which the scenario does not consider application-level caching.
As a consequence,  to support an arrival rate of $200$ queries/second considering  the same system configurations of Scenario~$4$, our model now indicates that a cluster composed of $300$ index servers ($3$~cluster replicas $\times$ $100$~index servers in a cluster, each replica supporting an arrival rate of $65$~queries/second and guaranteeing an average query system response time of $282$~milliseconds) would be able to achieve the desired performance, as it benefits from the performance improvements resulting from application-level caching.
%

\section{Conclusions}
\label{section:conclusions}


As the main contribution of this paper,
we proposed and investigated a methodology for analyzing the performance
of search engines. Based on the proposed methodology,
we developed a capacity planning model for vertical search engines that considers the imbalance in query service times among homogeneous index servers.
Our model, based on queueing theory, is simple and reasonably accurate. To fine tune the model, we ran experiments on
a small cluster of index servers. Once tuned, we compared the
predictions of our model with results we measured
empirically and found a high quality matching. Even at the saturation
point, the predictions of our model were reasonably accurate.

We illustrated how to apply our model to predict average query system response time
when adopting larger main memories,  faster CPUs and disks than those in use, as well as applying an application-level caching of query results at the broker.
We considered a realistic scenario, where a collection of $1$~billion documents is distributed over $100$~index servers.
In this scenario, we showed that the manager of a vertical search engine can quickly reach  predictions for upper bounds on the
average query system response time without having to run any live experiments.

Given the complexity of maintaining large scale search engines, and the simplicity
and reasonable accuracy of our model, we believe our model can be useful
in practice as it provides a conservative (upper bound) estimate on expected
performance. The achieved accuracy should be enough for capacity planning decisions.
As our model provides promising results, it may also be seen as a building block
for further research on capacity planning models for search engines that
may be extended to take into account other factors as discussed in the following.

A direction for future work is to extend our capacity planning model to support multiple processing threads at index servers.
Another direction for research is to improve our model to estimate the distribution function of the query system response time of a cluster of index servers.
This solution would be useful if the manager of the search engine requires the $q$-percentile of the expected query system response time to be less or equal than a given threshold.

Another direction for further research is to explicitly model the caching of query results---which allows the search engine to answer recently repeated queries at a very low cost since it is not necessary to process those queries---and caching of the inverted lists of query terms---which improves the query processing time for the new queries that include at least one term whose list is cached~\cite{Saraiva01}.
Identifying and analyzing the reasons for the use of disk cache, and eventually modeling the probability of disk cache hit, is also an interesting point to be addressed in future work.

Furthermore, future research may include a simulation-based analysis to verify the accuracy of our model predictions for larger clusters with thousands of index servers for supporting a collection with billions of documents, as illustrated in Section~\ref{section:applicability}.
Finally, another direction for further research is to develop an approach for finding the cost-optimal architecture for a vertical search engine, combining the strategies of collection partitioning and collection replication to satisfy operational requirements for query response time, query throughput, and server utilization.

\section{Acknowledgments}

This work was partially supported by the Brazilian National Institute of Science and Technology for the Web
(grant MCT/CNPq 573871/2008-6), Project InfoWeb (grant MCT/CNPq/CT-INFO 550874/2007-0), FAPES/MCT/CNPq grant 37711393/2007 (Claudine Badue); CNPq grant 30.0188/95-1 (Berthier Ribeiro-Neto); CNPq grant 301665/2008-7
(Artur Ziviani); and CNPq grant 30.5237/02-0 (Nivio Ziviani).

\bibliographystyle{plain}
\bibliography{paper}

\end{document}